\documentclass[aps,prb,preprint,amsmath,amssymb,aps,floats,epsfig]{revtex4}

\usepackage{graphicx}
\usepackage{dcolumn}
\usepackage{bm}
\usepackage{hyperref}
\usepackage{multirow}
\usepackage{array}
\usepackage{booktabs}
\usepackage{ctable}
\usepackage{upgreek}
\usepackage{epsfig}
\usepackage{mathrsfs}
\usepackage{amssymb}
\usepackage{amsbsy}
\usepackage{color}
\usepackage{cancel}
\usepackage{pifont}
\usepackage{marginnote}
\usepackage{float}
\newcommand{\bcen}{\begin{center}}
\newcommand{\ecen}{\end{center}}
\newcommand{\btab}{\begin{tabular}}
\newcommand{\etab}{\end{tabular}}
\newcommand{\bdes}{\begin{description}}
\newcommand{\edes}{\end{description}}

\newcommand{\beq}{\begin{equation}}
\newcommand{\eeq}{\end{equation}}
\newcommand{\bea}{\begin{eqnarray}}
\newcommand{\eea}{\end{eqnarray}}
\newcommand{\non}{\nonumber}

\newcommand{\bary}{\begin{array}}
\newcommand{\eary}{\end{array}}
\newcommand{\benum}{\begin{enumerate}}
\newcommand{\eenum}{\end{enumerate}}
\newcommand{\bi}{\bibitem}
\newcommand{\noi}{\noindent}
\newcommand{\bitem}{\begin{itemize}}
\newcommand{\eitem}{\end{itemize}}

\newcommand{\al}{\alpha}
\newcommand{\de}{\delta}

\newcommand{\ep}{\epsilon}
\newcommand{\ga}{\gamma}

\newcommand{\si}{\sigma}
\newcommand{\om}{\omega}
\newcommand{\dg}{\dagger}
\newcommand{\ua}{\uparrow}
\newcommand{\da}{\downarrow}
\makeatletter

\newcommand{\Rmnum}[1]{\expandafter\@slowromancap\romannumeral #1@}
\makeatother

\newcommand{\mylabel}[1]{\label{#1}} 

\begin{document}
\relax



\title{Effects of interactions on periodically driven dynamically localized 
systems}

\author{Adhip Agarwala$^1$ and Diptiman Sen$^2$}
\affiliation{\small{$^1$Department of Physics, Indian Institute of Science,
Bengaluru 560012, India \\
$^2$Centre for High Energy Physics, Indian Institute of Science, Bengaluru
560012, India}}



\date{\today}

\begin{abstract}
It is known that there are lattice models in which non-interacting 
particles get dynamically localized when periodic $\de$-function kicks are 
applied with a particular strength. We use both numerical and analytical
methods to study the effects of interactions in three different models in one
dimension. The systems we have considered 
include spinless fermions with interactions between nearest-neighbor sites, 
the Hubbard model of spin-1/2 fermions, and the Bose Hubbard model with 
on-site interactions. We derive effective Floquet Hamiltonians up to second 
order in the time period of kicking. Using these we show that interactions 
can give rise to a variety of interesting results such as two-body bound 
states in all three models and dispersionless few-particle bound states with 
more than two particles for spinless fermions and bosons. We substantiate 
these results by exact diagonalization and stroboscopic time evolution of 
systems with a few particles. 
We derive a pseudo-spin-1/2 limit of the Bose Hubbard system in the 
thermodynamic limit and show that a special case of this has an exponentially 
large number of degenerate eigenstates of the effective Hamiltonian. Finally we 
study the effect of 
changing the strength of the $\de$-function kicks slightly away from perfect 
dynamical localization; we find that a single particle remains dynamically 
localized for a long time after which it moves ballistically.
\end{abstract}

\maketitle

\section{Introduction}
\label{sec_intro}

Periodically driven quantum systems have been studied extensively for many 
years as they exhibit a wide variety of interesting phenomena. These include 
the coherent destruction of tunneling~\cite{grossmann91,kayanuma94}, the 
generation of defects~\cite{mukherjee08,mukherjee09}, dynamical 
freezing~\cite{das10}, dynamical saturation~\cite{russomanno12} and 
localization~\cite{alessio13,bukov14,nag14,nag15,agarwala16},
dynamical fidelity~\cite{sharma14}, edge singularity in the probability
distribution of work~\cite{russomanno15} and thermalization~\cite{lazarides14a}
(for a review see Ref.~\onlinecite{dutta15}). There have also been proposals 
of Floquet driving of graphene by radiation~\cite{gu11,kitagawa11,morell12,
sentef15}, Floquet topological insulators and the
generation of topologically protected edge states~\cite{kitagawa10,lindner11,
jiang11,trif12,gomez12,dora12,cayssol13,liu13,tong13,rudner13,katan13,
lindner13,kundu13,basti13,schmidt13,reynoso13,wu13,manisha13,perez1,perez2,
perez3,reichl14,manisha14}; some of these aspects have been experimentally
studied~\cite{kitagawa12,rechtsman13a,rechtsman13b,rechtsman14,tarruell12,
jotzu14}.

The effects of interactions between electrons in periodically driven systems 
have received much attention in recent times~\cite{Eckardt05,Rapp12,Zheng14,
Greschner14,lazarides14b,rigol14,ponte15a,lazarides14c,ponte15b,Eckardt15,
BukovPRL16,lazarides14d,keyser16,else16a,else16b}. It has been shown that 
a sinusoidal perturbation 
of the Hubbard model can lead to coherent destruction of tunneling, creation 
of gauge fields, and density-dependent tunneling~\cite{Itin15}. The effects 
of interactions on Floquet topological insulators have been examined in 
Ref.~\onlinecite{mikami16}. It has been shown that interactions can lead to a 
chaotic and topologically trivial phase in the periodically driven Kitaev 
model~\cite{su16}. 
The impact of such driving on the stability of a bosonic fractional Chern 
insulator has been investigated~\cite{Raciunas16}. Interestingly some of these 
systems have been realized experimentally demonstrating correlated hopping in 
the Bose Hubbard model~\cite{Meinert2016} and many-body 
localization~\cite{BordiaPRL16,Bordia16}, and realizing bound states for 
two particles in driven photonic systems~\cite{Mukherjee16}.

A particularly interesting phenomenon which can arise due to driving is 
dynamical localization. Here the particles become perfectly localized in 
space due to periodic driving of some parameter in the Hamiltonian. Examples 
of systems showing dynamical localization include driven two-level 
systems~\cite{grossmann91}, classical and quantum kicked 
rotors~\cite{chirikov81,fishman82,ammann98,tian11,nieuwenburg12}, the Kapitza 
pendulum~\cite{kapitza51,broer04}, and bosons in an optical 
lattice~\cite{horstmann07}. It has been shown that remnants of dynamical 
localization may survive even in the presence of strong disorder~\cite{Roy15}.

In this paper, we will study the effects of periodic driving on a number of
systems with interacting particles. The motivation for this is as follows.
Suppose we consider a system without any interactions and subject it to a
periodic driving which dynamically localizes the particles. This means
that the effective Floquet Hamiltonian of the system has no kinetic energy; 
for instance, in a tight-binding model, the effective hopping amplitude 
is zero. We now add interaction terms to the Hamiltonian. We may then expect
that the properties of the system will be entirely dominated by these terms.
Systems which are dominated by interactions often have interesting 
ground states, such as fractional quantum Hall systems and fractional 
Chern insulators in general~\cite{regnault11,sheng11,neupert11,wang11,tang11}.
We will therefore look at the effects of interactions on systems which are 
dynamically localized in the absence of interactions. We will consider only 
one-dimensional models here although many of our results can be easily
generalized to higher dimensions. As will become clear, new effective
hopping terms are generated by the interactions; these lead to dispersing
two-particle bound states and dispersionless bound states with more than
two particles. 
We will mainly study systems with a few particles rather than a finite density
of particles. However, for the Bose Hubbard model we will study the 
eigenstates of the effective Hamiltonian of a large system with a finite 
particle density in a particular limit.

The plan of our paper is as follows. In Sec.~\ref{sec_dl}, we will show that 
particles moving in a bipartite lattice with a non-interacting Hamiltonian 
can become dynamically localized if periodic $\de$-function kicks with a 
particular strength given by $\al = \pi$ are applied to the sublattice 
potential. (The advantage of looking at periodic $\de$-function kicks, in 
contrast to sinusoidal driving~\cite{nag15}, is that the problem can be studied
analytically to a large extent~\cite{nag14,dasgupta14,agarwala16}). The 
dynamical localization becomes clear when we view the system stroboscopically,
at intervals of time given by $2T$, where $T$ is the time period of the 
kicking. We find that the effective Hamiltonian which evolves the 
system for time $2T$ is exactly zero for this non-interacting problem.
In Sec.~\ref{sec_interact}, we will show how a generic model with interactions
can be studied by computing the effective Hamiltonian. This Hamiltonian can 
be derived as an expansion in powers of $T$, and we will carry out the 
expansion up to order $T^2$. In Sec.~\ref{sec_spinless}, we will consider a 
model of spinless fermions with nearest-neighbor interactions in one 
dimension. After deriving the effective Hamiltonian to order $T^2$, we will 
show that the system has two branches of two-body bound states; these states
move slowly if $T$ is small in appropriate units. We will also
show that there are bound states with three or more particles; these objects
have zero dispersion and do not move. We will demonstrate these results
both analytically and numerically. In Sec.~\ref{sec_spinhalf}, we will consider
the Hubbard model in one dimension, namely, a spin-1/2 model with on-site 
interactions. After deriving the effective Hamiltonian, we will show 
analytically and numerically that this has two branches of two-body bound 
states which are spin singlets. In Sec.~\ref{sec_bose}, we will study the Bose 
Hubbard model with on-site interactions in one dimension. We will derive the 
effective Hamiltonian and show that there are again two dispersing branches of
two-particle bound states and dispersionless bound states with more than
two particles. We will then consider a limit in which the interactions have 
a two-fold degenerate ground state on each site. After defining a
pseudo-spin-1/2 on each site, we derive an effective Hamiltonian for the
system. This contains both two-spin and three-spin interactions. For a special
case (one in which particle occupation numbers zero and 1 are degenerate on 
each site), we show that a class of degenerate eigenstates of the effective 
Hamiltonian 
can be found exactly and the number of such states grows exponentially with
the system size. In Sec.~\ref{sec_pert}, we will study the effects of two
kinds of perturbations on dynamical localization when there are no 
interactions. 
First, we study what happens if the strength of the $\de$-function kicks, 
$\al$, is slightly different from $\pi$. We show that a particle remains
dynamically localized for a long time which is of the order of $1/|\pi - \al|$.
After that time the particle begins to move ballistically with a 
maximum velocity which is of the order of $|\pi - \al|$. Second, we study
what happens if $\al = \pi$ but there is some randomness in the 
nearest-neighbor hoppings. In this case, we find that a particle remains
dynamically localized if we view at intervals of time $2T$. We end in 
Sec.~\ref{sec_concl} with a summary of our main results and some directions 
for future work.

\section{Dynamical Localization}
\label{sec_dl}

In this section we will consider a general non-interacting Hamiltonian on a 
bipartite lattice which respects the sublattice symmetry. We will show that 
such a system exhibits dynamical localization when periodic $\de$-function 
kicks with a particular strength are applied to the sublattice potential.

We consider a Hamiltonian on a bipartite lattice given by 
\beq H_{NI} ~=~ \sum_{ij} ~t_{ij} ~(c^{\dg}_{iA} c_{jB} + H. c.), \eeq
where $i$ and $j$ represent site labels residing on the two sublattices $A$ and 
$B$. We now apply periodic $\de$-function kicks to the sublattice potential 
as follows: the kicking part of the Hamiltonian, $H_K$, is given by
\beq H_K ~=~ \al ~\sum_{n=-\infty}^\infty \de (t-nT) ~\left( \sum_i 
n_{iA} ~-~ \sum_j n_{jB} \right), \eeq
where $n_{iA} =c^{\dg}_{iA} c_{iA}$ and $n_{jB} = c^{\dg}_{jB} c_{jB}$ 
denotes the number of particles on site $i$ on sublattice $A$ and site $j$
on sublattice $B$. We define the total number of particles on the two 
sublattices as
\bea N_A ~=~ \sum_i n_{iA} ~~~~{\rm and}~~~~ N_B ~=~ \sum_j n_{jB} . \eea
Without the kick the time evolution operator is given by 
\beq U_{NI} ~=~ e^{-iH_{NI} T}. \eeq 
(We will set $\hbar = 1$ in this paper).
The time evolution corresponding to the kick is
\beq U_K = e^{-i \al ~(N_A ~-~ N_B)}. \eeq
The total time evolution operator $U$ for a time period $T$ is the product 
of the two operators above. For $\al = \pi /2$, we obtain 
\bea U ~=~ U_K U_{NI} ~=~ e^{-\frac{i\pi}{2} ~(N_A ~-~ N_B)}~
e^{-iH_{NI} T}. \label{flo} \eea
Since the number operators of different sites commute, we can use the 
identities in Eqs.~\eqref{eq:expiden} and \eqref{eq:iden} to obtain
\bea U &=& e^{- \frac{i\pi}{2} N_A} ~\exp \left(-i T ~\sum_{ij} ~
t_{ij} ~(c^{\dg}_{iA} e^{-\frac{i\pi}{2}} c_{jB} + e^{\frac{i\pi}{2}} 
c^{\dg}_{jB}c_{iA} ) \right) ~e^{\frac{i\pi}{2} N_B } \non \\
&=& \exp \left( -iT \sum_{ij} ~t_{ij} ~(c^{\dg}_{iA} e^{-\frac{i\pi}{2}}
e^{-\frac{i\pi}{2}}c_{jB} + e^{\frac{i\pi}{2}} e^{\frac{i\pi}{2}} c^{\dg}_{jB}
c_{iA}) \right) ~e^{- \frac{i\pi}{2} (N_A - N_B)} \non \\
&=& e^{iH_{NI}T} ~e^{- \frac{i\pi}{2} (N_A - N_B)}. \eea
Hence the kick converts 
\beq H_{NI} \rightarrow -H_{NI}, \eeq
and the evolution operator for two time periods $2T$ is 
\bea U^2 &=& e^{- \frac{i\pi}{2} (N_A - N_B)}
~e^{-i H_{NI} T} ~e^{i H_{NI} T} ~e^{- \frac{i\pi}{2} (N_A - N_B)} \non \\
&=& e^{- i\pi (N_A - N_B)} \non \\
&=& e^{- i\pi (N_A + N_B)}, \label{u2t} \eea
where the last line follows from the previous line because $N_B$ is
an integer. Eq.~\eqref{u2t} implies that after time $2T$, all wave functions 
remain exactly the same up to a factor of $\pm 1$. Hence if we view the system
with any number of particles at intervals of $2T$, all the particles will 
appear to be localized. Note that this argument for dynamical localization 
works in exactly the same way for bosons, since the algebra leading up to 
Eq.~\eqref{u2t} remains the same.

Eq.~\eqref{u2t} shows that $U^2$ is equal to $I$ if the total number of 
particles $N_{tot} = N_A + N_B$ is even and $-I$ if $N_{tot}$ is odd. We can 
now define an effective Hamiltonian for evolution for time $2T$ as follows.
\bea U^2 &=& e^{-i 2T H_{eff}}, \non \\
{\rm implying}~~ H_{eff} &=& \frac{i}{2T} \ln (U^2). \eea
Since $U^2 = \pm I$, we see that
\bea H_{eff} &=& 0 ~~{\rm if}~~ N_{tot} ~~{\rm is ~even}, \non \\
&=& \frac{\pi}{2T} ~~{\rm if}~~ N_{tot} ~~{\rm is ~odd}. \eea
Hence, for a non-interacting problem, the effective Hamiltonian only depends 
on $N_{tot}$ and has no information about $H_{NI}$. 

We note that $H_{eff}$ and therefore its eigenvalues (called 
quasienergies) are only defined up to multiples of $\om = 2\pi /T$. In the 
following sections we will derive $H_{eff}$ as an expansion in powers of $T$ 
in the limit that $\om$ is much larger than all the other energy scales of 
the problem like the nearest-neighbor hopping amplitude $\ga$. This implies 
that the band width, which is
typically given by $4 \ga$, is much smaller than $\om$. Since $\om$ 
is much larger than the energy difference between any two states in the 
band, we will not need to consider the possibility of resonances.


The above analysis of dynamical localization by periodic $\de$-function 
kicks can be generalized as follows. Consider a kicking Hamiltonian 
\beq H_K ~=~ \sum_{n=-\infty}^\infty ~\de (t - nT) ~(\al N_A - \beta N_B). \eeq
where $\al + \beta = \pi$. The time evolution operator for one time period 
is now given by
\bea U &=& e^{-i (\al N_A - \beta N_B)} ~\exp \left({ -i \sum_{ij} t_{ij}
\{c^{\dg}_{iA} c_{jB} + c^{\dg}_{jB}c_{iA} \} T} \right) \non \\
&=& \exp \left({ -i \sum_{ij} t_{ij}\{c^{\dg}_{iA} c_{jB} e^{-i(\al+\beta)} 
+ c^{\dg}_{jB}c_{iA} e^{i(\al+\beta)} \} T} \right) ~e^{-i (\al N_A - \beta 
N_B)}. \eea
As we can see, this has the effect of converting $H_{NI} \to - H_{NI}$
for any $\al, ~\beta$ which satisfy $\al+\beta = \pi$. Therefore the 
evolution operator for time $2T$ is 
\bea U^2 &=& e^{-i2 (\al N_A - \beta N_B)} \non \\
&=& e^{-i 2 \al N_{tot}}, \eea
where we have used the facts that $\al + \beta = \pi$ and $N_B$ is an integer.
The effective Hamiltonian is now
\beq H_{eff} ~=~ \frac{\al}{T} ~N_{tot}. \eeq
Thus, by changing the values of $\al, ~\beta$ and the total number of 
particles $N_{tot}$ in the system, we can modulate the value of the
quasienergy (the eigenvalue of $H_{eff}$) at which dynamical localization 
occurs.

In the rest of this paper, we will take $\al = \pi, ~\beta =0$ so that the 
periodic $\de$-function kicks are applied to only the $A$ sublattice;
the kicking operator is therefore
\beq U_K ~=~ e^{-i \pi N_A}. \label{uk} \eeq
Then the eigenvalue of the non-interacting effective Hamiltonian will always 
be zero. This will allow us to look at the effects of interactions more 
cleanly.

\section{Interactions}
\label{sec_interact}

We will now consider what happens if we take the dynamically localized
system considered in the previous section and turn on density-density
interactions between the particles. We will first make some general
remarks before turning to three examples of interacting systems. In each 
case, we will use perturbation theory to calculate the effective Hamiltonian
for evolution by a time $2T$.

We consider a generic interaction term of the kind 
\beq H_I ~=~ U \sum_{ij} n_i n_j, \eeq
where $n_i$ denotes the particle number at site $i$. This term commutes with 
the kicking Hamiltonian $H_K$. Hence, when we pass the unitary operator 
$U_K = e^{-iH_K T}$ across the Hamiltonian $H_I + H_{NI}$, the sign of $H_I$ 
does not flip while the sign of $H_{NI}$ flips. 
The effective Hamiltonian after two time periods is therefore
\beq e^{-iH_{eff} 2 T} ~=~ e^{-i(-H_{NI} +H_I)T} ~e^{-i(H_{NI}+H_I)T}.
\mylabel{effHam} \eeq

Now we use Eqs.~\eqref{eq:BCH} and \eqref{eq:BCHSD} to evaluate the above 
term. Setting $C=-iH_I T$ and $D=iH_{NI}T$ in those equations, we obtain
\beq -iH_{eff} 2T ~=~ -i 2H_I T ~+~ [H_{NI}, H_I]T^2 ~+~ \frac{i}{3} \left( 
\left[H_I,H_{NI}\right] H_{NI} ~+~ H_{NI} \left[H_{NI},H_I \right] \right) 
T^3 ~+~ \cdots. \eeq
This implies that
\beq H_{eff} ~=~ H_I ~+~ \frac{iT}{2} ~[H_{NI}, H_I] ~-~ \frac{T^2}{6}
[H_{NI},[H_{NI}, H_I]] ~+~ \cdots. \mylabel{EffH} \eeq
This equation is one of the central results of this work. It provides a 
perturbative expansion if we assume that $T$ is a small parameter. 

We now prove another result which will be important in our analysis later.
Namely, $H_{eff}$ only contains odd powers of $H_I$. This can be proved as 
follows. Let 
\beq \ln (e^{C+D}e^{C-D}) = f(C,D). \eeq
Then
\bea f(-C, D) &=& \ln (e^{-C+D}e^{-C-D}) \non \\
&=& \ln \left((e^{C-D})^{-1}(e^{C+D})^{-1}\right) \non \\
&=& - \ln (e^{C+D} e^{C-D}) \non \\
&=& -f(C,D). \label{odd} \eea
This implies that $f(C,D)$ is an odd function of $C$. Now we recall that $C$ 
is proportional to $H_I$. This shows that $H_{eff}$ only contains odd 
powers of $H_I$. 

To get an idea of the kinds of terms that can arise due to the commutators
in Eq.~\eqref{EffH}, we consider a particular interaction term given by
\beq H^{ij}_I ~=~ n_i n_j \eeq
where $i \ne j$, and a hopping term given by
\beq H^{kl}_{NI} =c^\dg_k c_l + c^\dg_l c_k \eeq
where $k \ne l$. We now look at the commutator of these interacting and 
non-interacting terms. We find the following.

\beq
\begin{array}{|c|c|c|}
\hline
i \neq k,l & j \neq k, l & [H^{kl}_{NI},H^{ij}_I] = 0 \\
i = k(l) & j = l(k) & [H^{kl}_{NI},H^{ij}_I] = 0 \\
i = k & l \neq j & [H^{kl}_{NI},H^{kj}_I] = n_j (-c^{\dg}_kc_l + 
c^\dg_l c_k) \\
\hline
\end{array} \eeq
\vspace*{.1cm}

We note the interesting fact that the commutator with interactions leads to 
correlated hoppings where the hopping is proportional to the particle number 
at some site. In the next few sections we will look at some well-known 
interacting models in one dimension systems and find the effective Hamiltonian
that is generated by periodic $\de$-function kicks. The commutator 
manipulations were partly performed using Ref.~\onlinecite{zitko11}.

Before ending this section, we note that when the driving frequency 
$\om = 2\pi/T$ is large, a Floquet-Magnus expansion in powers of $1/\om$ can 
be used to find the effective Floquet Hamiltonian~\cite{bukov14,mikami16}. 
This works well when the time-dependent Hamiltonian $H(t)$ has only a few 
harmonics, namely, when only a few terms are non-zero in
\beq H(t) ~=~ \sum_{n=-\infty}^\infty ~H_n ~e^{-in \om t}. \label{hn} \eeq
For instance, if only $H_0, ~H_1$ and $H_{-1}$ are present in Eq.~\eqref{hn},
we get
\beq H_{eff} ~=~ H_0 ~+~ \frac{[H_{-1}, H_1]}{\om}. \eeq
However, in the case of periodic $\de$-function kicks, an infinite number of
terms are present in \eqref{hn} and the Floquet-Magnus expansion is not 
convenient.

\section{Spinless fermions with nearest-neighbor interactions}
\label{sec_spinless}

In this section, we will consider a system of spinless fermions hopping on 
a one-dimensional chain with nearest-neighbor interactions and periodic 
boundary conditions. Given $N$ sites we have $2^N$ states which are labeled 
by the occupancies, zero or 1, of the different sites. The Hamiltonian is
\beq H ~=~ \sum_{j=1}^N ~[- \ga (c_j^\dg c_{j+1} + H. c.) ~+~ V n_j n_{j+1}], 
\label{ham1} \eeq
with $c_{N+1} \equiv c_1$. Note that 
the Hamiltonian does not mix the various sectors of total particle number 
$N_{tot} = \sum_{j=1}^N c_j^\dg c_j$. Hence we can consider a state with 
a given number of particles and look at its time evolution. For the sector 
with $p$ particles the number of relevant states is given by ${}^N\!C_p$. 
In the absence of kicking, this model is exactly solvable by the Bethe
ansatz and all its energy levels are known for any number of 
particles~\cite{mattis93,sutherland04}.

Following the notation in the previous section we identify
\bea H_{NI} &=& - ~\ga ~\sum_{j=1}^N ~(c_j^\dg c_{j+1} + H. c.), \non \\ 
H_I &=& V ~\sum_{j=1}^N ~n_j n_{j+1}. \eea
We now evaluate $[H_{NI}, H_I]$. The relevant terms are of the kind
\bea && [c_j^\dg c_{j+1} + c^\dg_{j+1}c_j , n_{j-1} n_j + n_j n_{j+1} + 
n_{j+1} n_{j+2} ] \non \\
&& =~ (c^\dg_{j+1}c_j -c_j^\dg c_{j+1}) (n_{j-1} - n_{j+2}). \eea
Next, we evaluate $[H_{NI},[H_{NI},H_I]]$ which involves terms like
\beq -\ga V [ H_{NI},(c^\dg_{j+1}c_j -c_j^\dg c_{j+1}) (n_{j-1} - n_{j+2}) ].
\eeq
This gives 
\bea && \ga^2 V ~[ 2(n_j - n_{j+1}) ~(n_{j-1}-n_{j+2}) \non \\
&& ~~+~ (c^\dg_{j-1}c_{j+1} + c^\dg_{j+1}c_{j-1}) ~(n_{j+2}-n_j) ~+~
(c^\dg_j c_{j+2} + c^\dg_{j+2}c_j) ~(n_{j-1}-n_{j+1}) \non \\
&& ~~+~ (c^\dg_j c_{j+1} - c^\dg_{j+1} c_j) ~(c^\dg_{j-1} c_{j-2} - c^\dg_{j-2}
c_{j-1}) ~+~ (c^\dg_{j+1}c_j -c^\dg_j c_{j+1}) ~(c^\dg_{j+2}c_{j+3} - 
c^\dg_{j+3} c_{j+2})]. \non \\
&& \eea

Using Eq.~\eqref{EffH}, we see that the total effective Hamiltonian up to 
terms of order $\ga^2 T^2$ (this is a dimensionless parameter) is given by
\bea H_{eff} &=& V ~\sum_j n_j n_{j+1} ~-~\frac{i \ga T V}{2} ~\sum_j ~
(c^\dg_{j+1}c_j -c_j^\dg c_{j+1})~(n_{j-1} - n_{j+2}) \non \\
&& - \frac{\ga^2 T^2 V}{3} ~\sum_j \Big( (n_j - n_{j+1})(n_{j-1} - n_{j+2}) 
\non \\ 
&& ~~~~~~~~~~~~~~~~~~~+~ \frac{1}{2} ~(c^\dg_{j-1}c_{j+1} + c^\dg_{j+1}
c_{j-1}) ~(n_{j+2} + n_{j-2}- 2 n_j) \non \\
&& ~~~~~~~~~~~~~~~~~~~-~ (c^\dg_{j-2}c_{j-1} - c^\dg_{j-1}c_{j-2}) ~(c^\dg_j
c_{j+1} -c^\dg_{j+1} c_j) \Big). \mylabel{eq:effHfer} \eea

It is interesting to note the scales of the various terms in 
Eq.~\eqref{eq:effHfer}. We see that the first three terms in the effective 
Hamiltonian all have the same energy scale as $V$, and $\ga T$ is the only 
tuning parameter. From the result we had proved using Eq.~\eqref{odd}, we know 
that the next higher order terms will be of order $\ga^3 T^3 V$ and 
$\ga T^3 V^3$.

For a system with only one particle located at, say, site $j$, it is clear 
from Eq.~\eqref{eq:effHfer} that the hopping amplitude to any other site
is zero, regardless of the value of $V$. This is expected since
interactions only play a role if there are at least two particles.

\subsection{Two-particle bound states}

We can use the Hamiltonian in Eq.~\eqref{eq:effHfer} to find eigenstates of a 
system with two or more particles. In particular, we can look for bound states
in which the wave function goes to zero when one or more of the particles
goes far away from the other particles. For example consider the case of two 
particles. We look for a bound state solution of the form
\beq |\psi_k \rangle ~=~ \sum_j ~[a e^{ik(j+1/2)}|j,j+1 \rangle ~+~ 
b e^{ik(j+1)} |j, j+2 \rangle], \mylabel{eq:varans} \eeq
where $a, ~b$ are some complex numbers that we have to determine while 
$k$ represents the center-of-mass momentum. For periodic boundary conditions,
we must have $k= 2\pi m/N$, where $m = 0,1,\cdots,N-1$. 

We now want to solve the eigenvalue equations 
\beq H_{eff} |\psi_k \rangle ~=~ E |\psi_k \rangle. \label{eq:eig} \eeq
To do this, we first look at the effect of each of the terms
in the Hamiltonian in Eq.~\eqref{eq:effHfer} on the two parts of the wave 
function in Eq.~\eqref{eq:varans}. This is shown in Tables I and II; 
a sum over $j$ from $1$ to $N$ is assumed in those tables.

\begin{table}[H]
\begin{center}
\begin{tabular}{|c|c|c|}
\hline 
Terms in $H_{eff}$ & Acting on $ae^{ik(j+1/2)}|j,j+1\rangle$ \\ 
\hline 
$V n_j n_{j+1}$ & $Vae^{ik(j+1/2)}|j,j+1\rangle$ \\ 
\hline 
$- \frac{i\ga TV}{2} (c^\dg_{j+1}c_j -c_j^\dg c_{j+1}) (n_{j-1} - n_{j+2})$ & 
$- \frac{i\ga TV}{2} ae^{ik(j+1/2)} (|j,j+2\rangle + |j-1,j+1\rangle)$ \\ 
\hline 
$- \frac{\ga^2 T^2 V}{3} (n_j - n_{j+1})(n_{j-1}-n_{j+2})$ & $- \frac{2\ga^2
T^2 V}{3} a e^{ik(j+1/2)}|j,j+1\rangle$ \\ 
\hline 
$- \frac{\ga^2 T^2 V}{6} (c^\dg_{j-1}c_{j+1} + c^\dg_{j+1}c_{j-1}) 
(n_{j+2} + n_{j-2}- 2 n_j)$ & $- \frac{\ga^2 T^2 V}{6} a
e^{ik(j+1/2)} (2|j-1,j\rangle + 2|j+1,j+2\rangle$ \\
& $+ |j,j+3\rangle + |j-2,j+1\rangle)$ \\
\hline 
$- \frac{\ga^2 T^2 V}{3} (c^\dg_{j-1}c_{j-2} - c^\dg_{j-2}c_{j-1}) (c^\dg_j
c_{j+1} -c^\dg_{j+1}c_j)$ & $- \frac{\ga^2 T^2 V}{3} ae^{ik(j+1/2)} 
|j-1,j+2 \rangle$ \\ 
\hline
\hline 
\end{tabular} 
\end{center}
\caption{Effect of various terms in $H_{eff}$ acting on the first term in 
$|\psi_k \rangle$.} 
\end{table}

\begin{table}[H]
\begin{center}
\begin{tabular}{|c|c|c|}
\hline 
Terms in $H_{eff}$ & Acting on $be^{ik(j+1)}|j,j+2\rangle$ \\ 
\hline 
$V n_j n_{j+1}$ & $zero$ \\ 
\hline 
$- \frac{i\ga TV}{2} (c^\dg_{j+1}c_j -c_j^\dg c_{j+1}) (n_{j-1} - n_{j+2})$ & 
$-\frac{i\ga TV}{2} be^{ik(j+1)} (|j,j+1\rangle + |j+1,j+2 \rangle)$ \\ 
\hline 
$- \frac{\ga^2 T^2 V}{3} (n_j - n_{j+1})(n_{j-1}-n_{j+2})$ & $\frac{2\ga^2 T^2
V}{3} b e^{ik(j+1)}|j,j+2\rangle$ \\ 
\hline 
$- \frac{\ga^2 T^2 V}{6} (c^\dg_{j-1}c_{j+1} + c^\dg_{j+1}c_{j-1}) 
(n_{j+2} + n_{j-2}- 2 n_j)$ & $zero$ \\ 
\hline 
$- \frac{\ga^2 T^2 V}{3} (c^\dg_{j-1}c_{j-2} - c^\dg_{j-2}c_{j-1}) (c^\dg_j
c_{j+1} -c^\dg_{j+1}c_j)$ & $\frac{\ga^2 T^2 V}{3} be^{ik(j+1)}(|j-1,j+1
\rangle+ |j+1,j+3\rangle)$ \\ 
\hline 
\end{tabular} 
\end{center}
\caption{Effect of various terms in $H_{eff}$ acting on the second term in 
$|\psi_k \rangle$.} 
\end{table}

By inspection, we see that a particular solution of Eq.~\eqref{eq:eig} is
given by $b=0$, $k = \pi$ and $E=V$; the corresponding wave function is 
\beq | \psi_k \rangle ~=~ \sum_j ~(-1)^j |j,j+1 \rangle. \label{psipi} \eeq
Note that this is an exact eigenstate of the Hamiltonian in 
Eq.~\eqref{ham1}; a state like this is called a singular solution of the Bethe
ansatz~\cite{nepomechie13,giri15}. In fact, the state in Eq.~\eqref{psipi} is 
an exact eigenstate of the kicking problem. This is because the number of
particles on sublattice $A$ is given by $N_A = 1$; hence this state is an 
eigenstate with eigenvalue $-1$ of the kicking operator $U_K$ in 
Eq.~\eqref{uk}.

We will now look for solutions of Eq.~\eqref{eq:eig} with arbitrary values of 
$k$ based on the terms of order $\ga^2 T^2 V$ coming from Tables I and II.
To do this consistently, we have to keep both the terms of order $\ga^2 T^2 V$
as they are and add the effect of the terms of order $\ga T V$ to second
order in perturbation theory, taking the first term in Eq.~\eqref{eq:effHfer},
$V n_j n_{j+1}$, as the unperturbed Hamiltonian.

{}From Table I, we find that the term of order $\ga T V$ takes an initial state
$|j,j+1\rangle$ with amplitude $ae^{ik(j+1/2)}$ to an intermediate state 
$|j,j+2\rangle$ and then back to the state $|j,j+1\rangle$. The numerator 
of this second order process is given by
\bea && \frac{\ga^2 T^2 V^2}{4} ~ae^{ik(j+1/2)}(1+e^{ik})~(|j+1,j+2\rangle 
+ |j,j+1 \rangle) \non \\
&=& \frac{\ga^2 T^2 V^2}{4} ~ae^{ik(j+1/2)}(1+e^{ik})(e^{-ik}+1) ~|j,j+1
\rangle. \eea
Dividing this by the energy denominator which is the difference of the 
unperturbed energies of the initial state $|j,j+1\rangle$ and the intermediate 
state $|j,j+2\rangle$, namely, $V-0 = V$, we obtain a contribution equal to
\beq \frac{\ga^2 T^2 V}{2} ae^{ik(j+1/2)}(1+\cos k) ~|j,j+1\rangle. \eeq 

Next we see from Table I that the three terms of order $\ga^2 T^2 V$ acting
on the state $|j,j+1 \rangle$ gives 
\beq - \frac{2 \ga^2 T^2 V}{3} (1 + \cos k)~ |j, j+1 \rangle, \eeq
where we have used the fact that $j$ is summed over, and we have ignored 
states which are not of the form $|j, j+1 \rangle$.

The total contribution is therefore
\bea && \Big(V + \ga^2 T^2 V (\frac{1}{2}-\frac{2}{3})(1+\cos k)\Big) ~a
e^{ik(j+1/2)} ~|j,j+1\rangle \non \\
&=& \Big(V - \frac{\ga^2 T^2 V}{3}\cos^2 \left(\frac{k}{2} \right) \Big) ~
ae^{ik(j+1/2)} ~|j,j+1\rangle. \eea

Similarly, from Table II we find that the term of order $\ga T V$ takes
an initial state $|j,j+2\rangle$ with amplitude $b e^{i (j+1)k}$ to an
intermediate state $|j,j+1\rangle$ and then back to the state $|j,j+2\rangle$.
The numerator of this second order process is
\bea && \frac{\ga^2 T^2 V^2}{4} be^{ik(j+1)} (1+e^{-ik}) ~(|j,j+2\rangle + 
|j-1,j+1 \rangle) \non \\
&=&\frac{\ga^2 T^2 V^2}{4} be^{ik(j+1)} (1+e^{ik})(1+e^{-ik}) ~|j,j+2
\rangle. \eea
The denominator is the difference of the unperturbed energies of the states
$|j,j+2\rangle$ and $|j,j+1\rangle$, namely, $0 - V = -V$. We therefore
find the contribution from this process to be
\beq -\frac{\ga^2 T^2 V}{2} be^{ik(j+1)}(1+\cos k) ~|j,j+2\rangle. \eeq
The total contribution is therefore 
\bea && \ga^2 T^2 V (-\frac{1}{2}+\frac{2}{3})(1+\cos k) be^{ik(j+1)} ~|j,j+2 
\rangle \non \\
&=& \frac{\ga^2 T^2 V}{3}\cos^2 \left(\frac{k}{2} \right) be^{ik(j+1)} ~
|j,j+2\rangle. \eea

Thus we find two branches of bound states: one branch has the dispersion
\beq E_{1k} ~=~ V - \frac{\ga^2 T^2 V}{3} \cos^2 \left(\frac{k}{2} \right), 
\label{e1k1} \eeq
in which the wave function has a large component in states of the form 
$|j,j+1\rangle$ and a small component in the states 
$|j,j+2 \rangle$, and the other branch has the dispersion
\beq E_{2k} ~=~ \frac{\ga^2 T^2 V}{3} \cos^2 \left(\frac{k}{2} \right), 
\label{e2k1} \eeq
in which the wave function is large for the states $|j,j+2 \rangle$ and small
for the states $|j,j+1\rangle$. We note that in both cases, the
group velocity is given by $v_g = |dE_{ak}/dk| = (\ga^2 T^2 V/6) |\sin k|$.
Hence these bound states move slowly if $\ga T$ is small.

We find that these are the only two-particle
bound states. All other two-particle states have a distance of three or more 
lattice spacings between the two particles, and all such states are completely
localized and have zero quasienergy. We have verified these results 
numerically. In Fig.~\ref{fig:eff1}
we compare the numerically obtained Floquet eigenvalues of a two-particle
system with the analytical expressions given in Eqs.~(\ref{e1k1}-\ref{e2k1})
for $V=1, ~T= 0.5$, and $\ga =1$. The agreement is seen to be extremely good.

\begin{figure}[H]
\centering
\includegraphics[width=14cm]{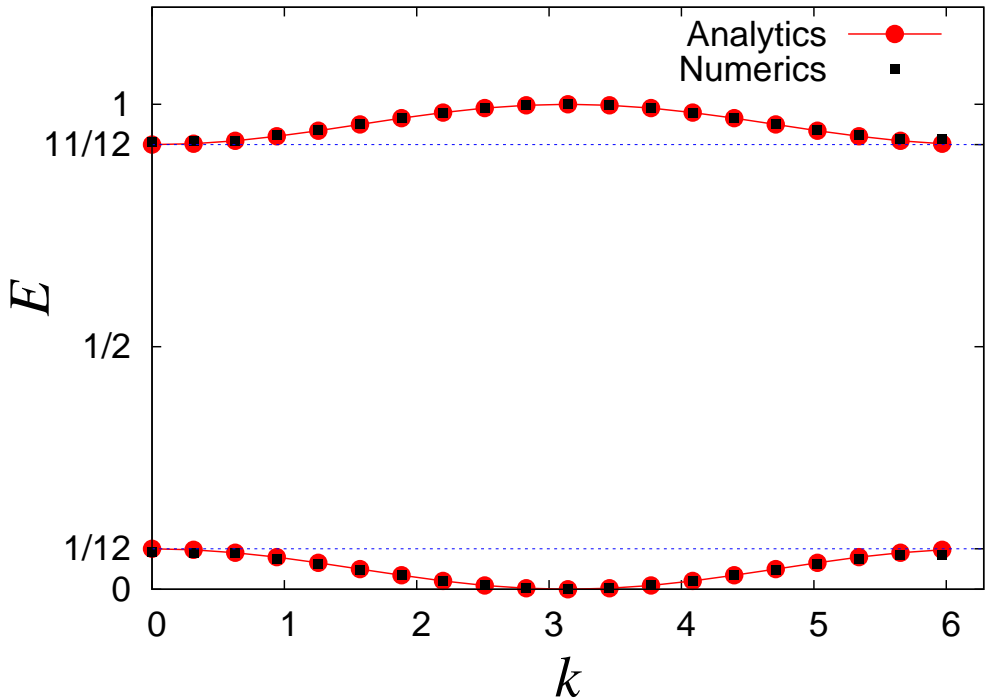}
\caption{Numerically obtained eigenvalues of the Floquet operator as compared 
with the analytical expressions in Eqs.~(\ref{e1k1}-\ref{e2k1}), for $V=1, ~T=
0.5$, and $\ga =1$. All other eigenvalues are zero. We have two particles on
20 sites.} \mylabel{fig:eff1} \end{figure}

In Figs.~\ref{fig:dyn1}-\ref{fig:dyn2}, we show the time evolution of two 
particles placed on a ring of 20 sites, with various initial conditions, 
interaction strengths and kicking; this system has 190 states. The time 
evolution is found by numerically evaluating the Floquet operator $U$ given 
in Eq.~\eqref{flo}; we have taken $\ga = 1$ and $T=0.5$ in all these studies.
We discuss below our numerical results and how they compare with what we 
expect from the effective Hamiltonian up to order $\ga^2 T^2 V$ that we 
have derived above. 

In Fig.~\ref{fig:dyn1}, we consider the time evolution when 
the initial state has the two particles on adjacent sites. The first two 
rows of this figure show that the particles spread out over the ring 
if there is no kicking; there is no major difference between 
the interacting and non-interacting cases. The third row shows that the 
particles are dynamically localized if there is kicking but no interaction. 
The fourth row shows that there is no dynamical localization if 
there is both kicking and interaction; however, since 
$\ga T = 0.5$ is small, the two particle bound state dispersion is almost 
flat which implies that the group velocity is small. Hence the particles 
spread out over the ring more slowly compared to the first two rows where 
there is no kicking. (In the fourth row, the eigenstates have large components 
on states of the form $|j,j+1 \rangle$). 

In Figs.~\ref{fig:dyn2}, we show the time evolution of two particles on 20 
sites in the presence of kicking. In Figs.~\ref{fig:dyn2} (i)-(ii), the
initial state has two particles which are separated by two lattice spacings. 
Figure (i) shows dynamical localization in the absence of interactions 
($V=0$). The behavior in Fig.~\ref{fig:dyn2} (ii) (where interactions are 
present with $V=1$) is similar to that in Fig.~\ref{fig:dyn1} (iv), except 
that the eigenstates now have large components on states of the form $|j,j+2 
\rangle$. In Figs.~\ref{fig:dyn2} (iii) and (iv), the initial state has two 
particles which are separated by three and four lattice spacings, namely, 
states of the form $|j,j+3 \rangle$ and $|j,j+4 \rangle$ respectively.
In these cases, the states has no overlap with the two-particle bound states
and therefore do not disperse. In the presence of interactions the particles 
seem to be localized. Looking more closely, we find that the particles
do spread a little bit when they are initially separated by three lattice 
spacings but not for four lattice spacings. This occurs because the wave 
function in the case of three lattice spacings has a small overlap with 
the two-particle bound states when we go to terms in the effective 
Hamiltonian which are of higher order than $\ga^2 T^2 V$.

\begin{figure}[H]
\centering
\includegraphics[width=14cm]{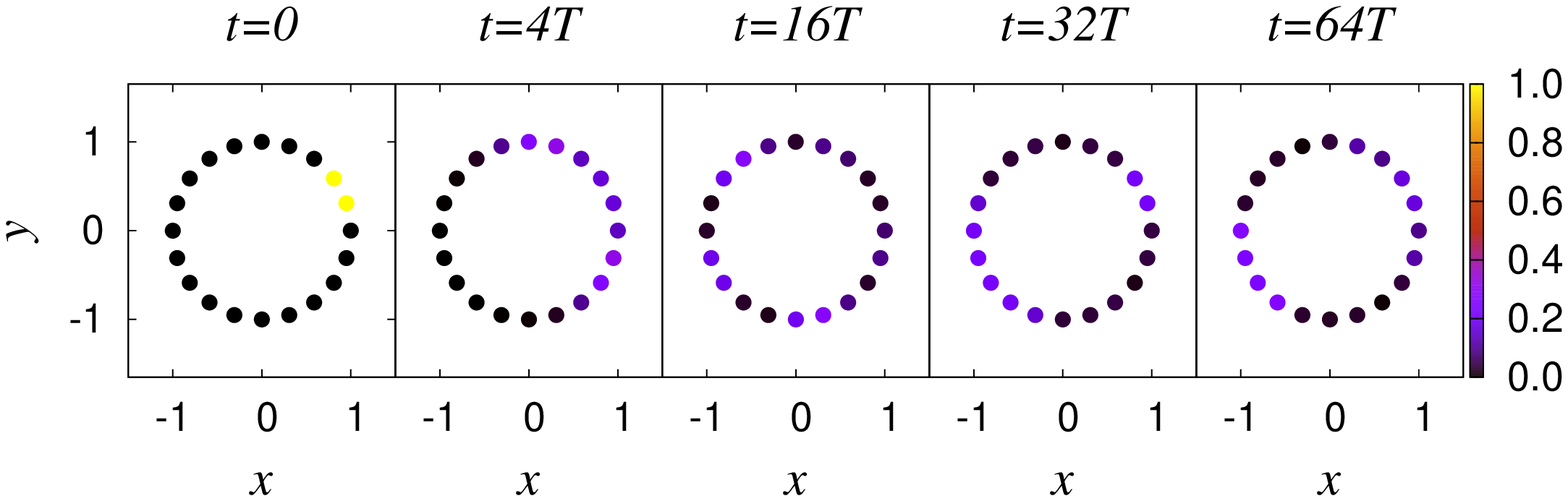}
\includegraphics[width=14cm]{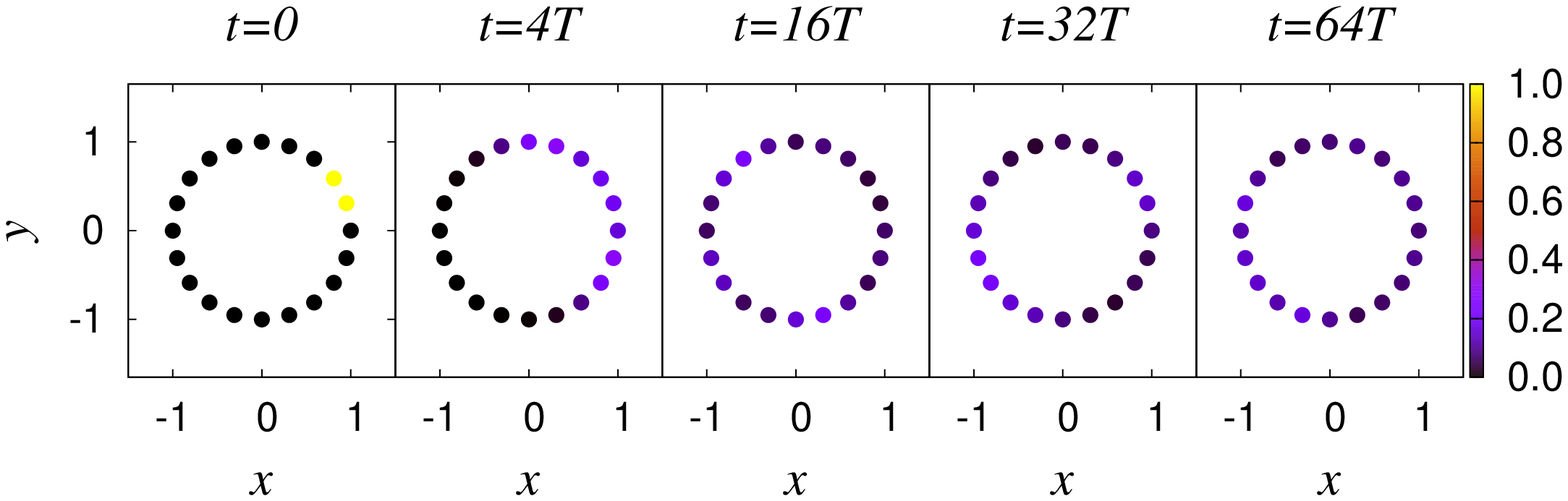}
\includegraphics[width=14cm]{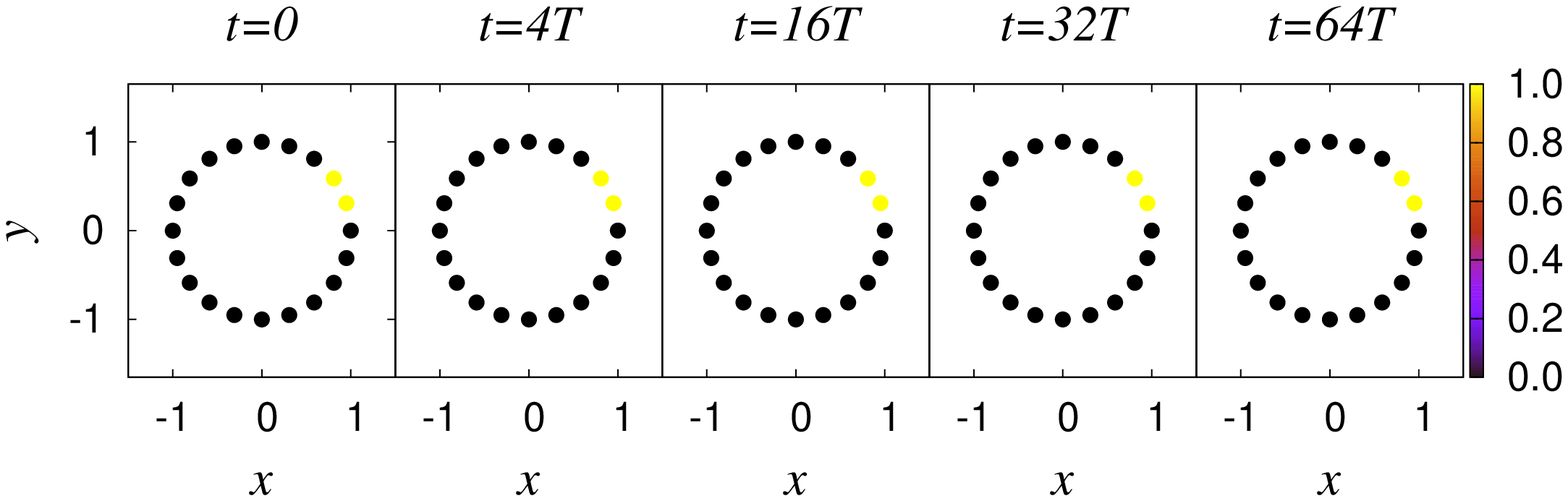}
\includegraphics[width=14cm]{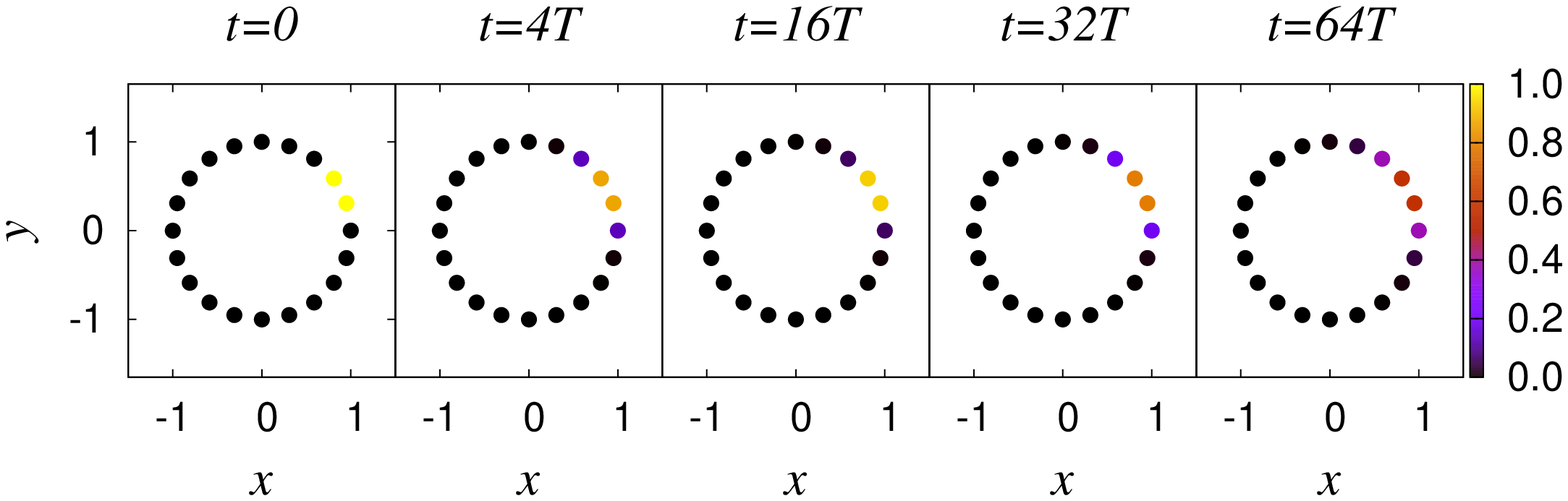}
\caption{Time evolution of a two-particle state for four cases: (i) $V=0$, 
no kicking, (ii) $V=1$, no kicking, (iii) $V=0$, with kicking, and (iv) $V=1$,
with kicking. In all cases $\ga = 1$ and $T =0.5$. There are two particles 
on $20$ sites, and they are initially located at two adjacent sites. The 
color shows the expectation value of the particle number at different sites.} 
\mylabel{fig:dyn1} \end{figure}

\begin{figure}[H]
\centering
\includegraphics[width=13.7cm]{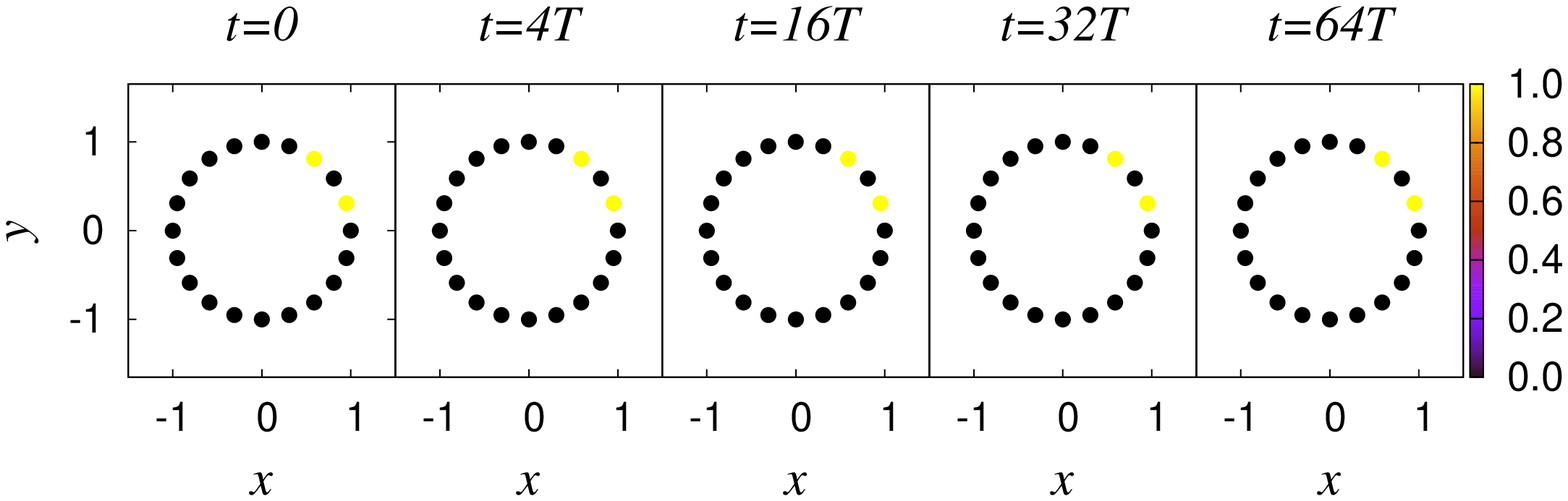}
\includegraphics[width=13.7cm]{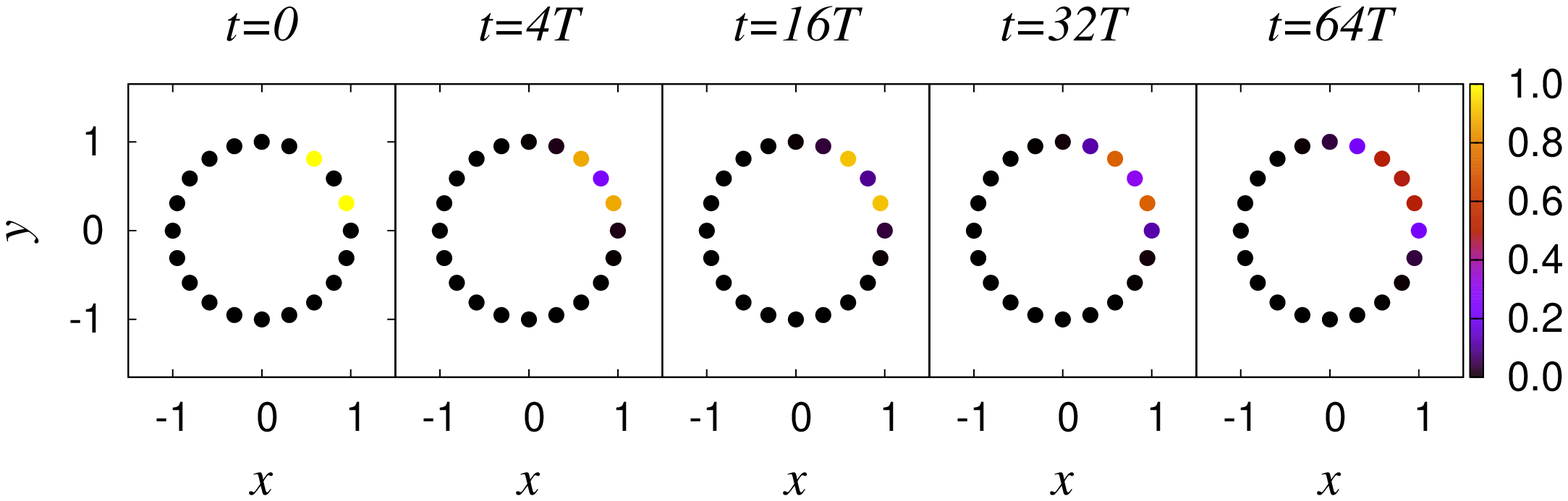}
\includegraphics[width=13.7cm]{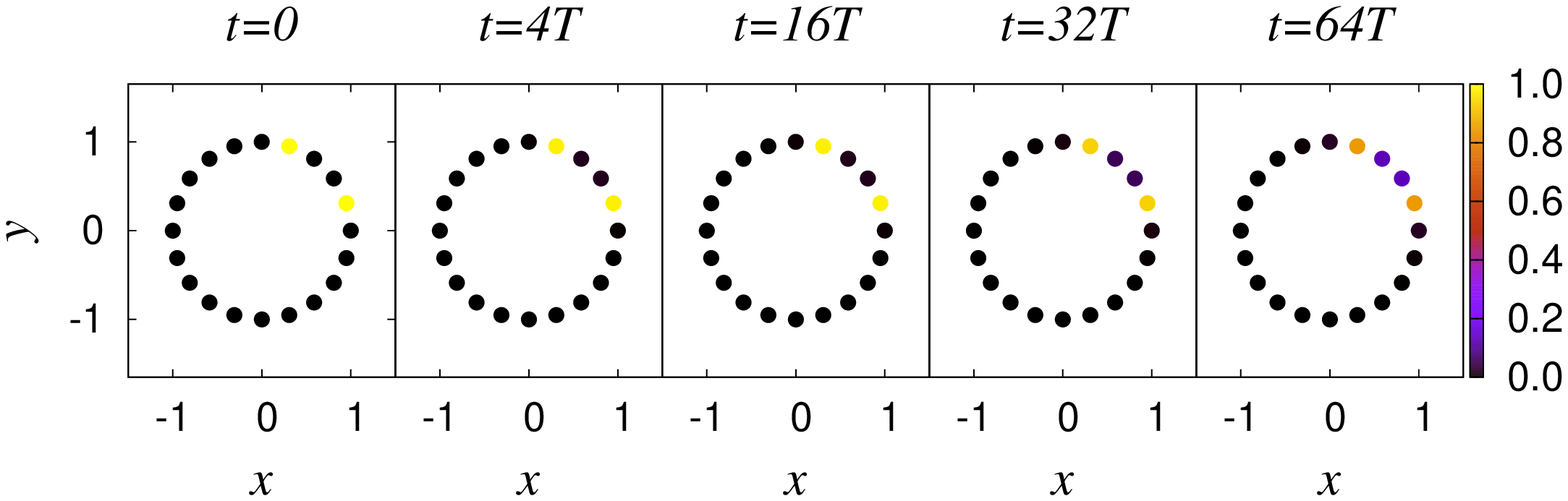}
\includegraphics[width=13.7cm]{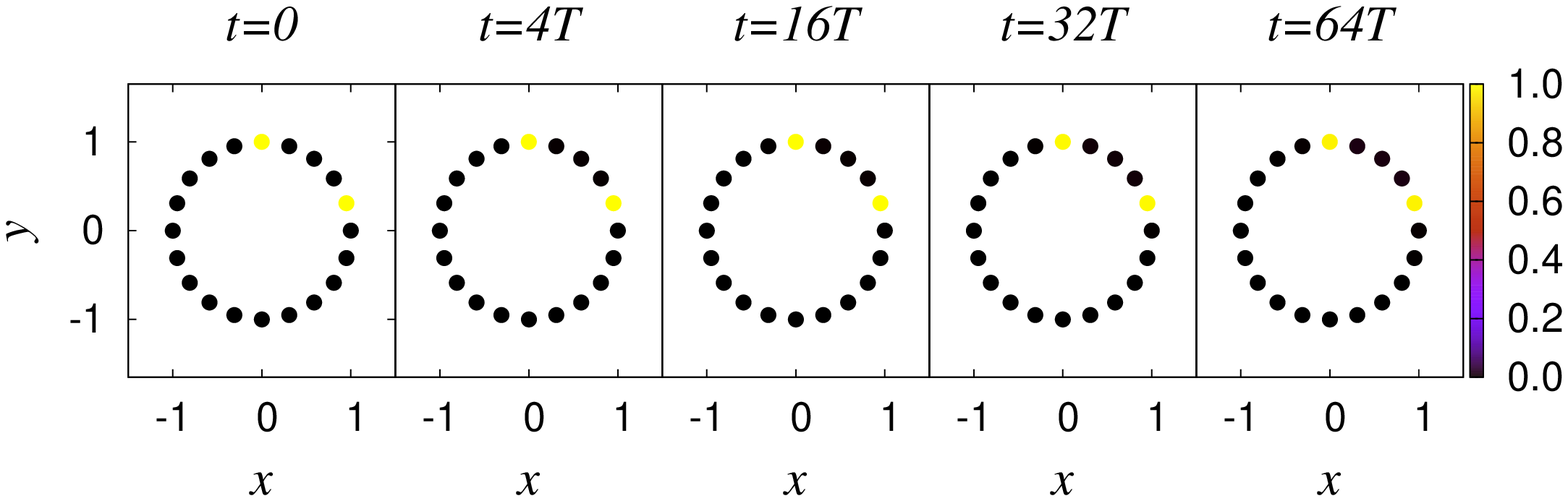}
\caption{Time evolution of a state with two particles on 20 sites in the 
presence of kicking, for four cases. In (i) the two particles are initially on
adjacent sites and there is no interaction ($V=0$). The state is dynamically 
localized due to kicking. In (ii)-(iv) the initial 
distance between the particles is progressively increased from two to four 
lattice spacings, and interactions are present with $V=1$. The color shows the 
expectation value of the particle number at different sites. Note that with 
increasing initial spacing the overlap with the two-particle bound states 
gets reduced, and the states get more localized.} \mylabel{fig:dyn2} 
\end{figure}

%

\subsection{States with three or more particles}

We will now study what happens when there are more than two particles.
We begin with the case of three particles. Assuming that 
they are on three neighboring sites, Table III shows the action of the 
different terms in Eq.~\eqref{eq:effHfer} on the state $|j,j+1,j+2 \rangle$.

\begin{table}[H]
\begin{center}
\begin{tabular}{|c|c|c|}
\hline 
Terms in $H_{eff}$ & Acting on $|j,j+1,j+2\rangle$ \\ 
\hline 
$V n_j n_{j+1}$ & $2V|j,j+1, j+2\rangle$ \\ 
\hline 
$- \frac{i\ga T V}{2} (c^\dg_{j+1}c_j -c_j^\dg c_{j+1}) (n_{j-1} - n_{j+2})$ 
& $- \frac{i\ga T V}{2} (|j-1,j+1,j+2\rangle$ \\ & ~~~~
$+ |j,j+1,j+3 \rangle)$ \\ 
\hline 
$- \frac{\ga^2 T^2 V}{3} (n_j - n_{j+1})(n_{j-1}-n_{j+2})$ & $- \frac{2\ga^2
T^2 V}{3} |j,j+1,j+2\rangle$ \\ 
\hline 
$- \frac{\ga^2 T^2 V}{6} (c^\dg_{j-1}c_{j+1} + c^\dg_{j+1}c_{j-1}) 
(n_{j+2} + n_{j-2}- 2 n_j)$ & $-$ \\ 
$- \frac{\ga^2 T^2 V}{3} (c^\dg_{j-1}c_{j-2} - c^\dg_{j-2}c_{j-1}) (c^\dg_j
c_{j+1} -c^\dg_{j+1}c_j)$ & \\ 
\hline 
\end{tabular}
\end{center}
\caption{Effect of various terms in $H_{eff}$ acting on the state $|j,j+1,j+2 
\rangle$. The $-$ symbol in the right column means we have states which only
contribute to the bound state at orders higher than $\ga^2 T^2 V$.}
\end{table}

A second order process involving the second term in $H_{eff}$ brings an initial
state $|j,j+1,j+2 \rangle$ back to itself, with an amplitude $\frac{\ga^2 T^2 
V^2}{4(2V-V)} = \ga^2 T^2 V/4$; the denominator $2V-V$ is the
difference in the unperturbed energies of the initial state and the 
intermediate states given by $|j-1,j+1,j+2 \rangle$ and $|j,j+1,j+3 \rangle$.
This process can happen in two ways since there are two possible intermediate 
states; hence this contribution is equal to $\ga^2 T^2 V /2$.
The total contribution is therefore, $(\frac{1}{2}-\frac{2}{3})\ga^2 T^2 V = -
\ga^2 T^2 V/6$. Therefore, we find non-dispersing eigenstates with quasienergy 
$2V- (\ga^2 T^2 V/6)$. The number of such states is equal to the number of 
sites $N$, since the index $j$ of the first particle can take any value from
1 to $N$.

In fact, there is an interesting solution for any number of particles $n$,
where $N-2 > n > 2$. Consider a state where $n$ particles are located next to 
each other. Due to the second order process described above, this is an 
eigenstate of $H_{eff}$ with quasienergy 
\beq E_n ~=~ (n-1)V ~-~ \frac{\ga^2 T^2 V}{6}. \label{en} \eeq
Thus we have non-dispersing states of clustered particles; the number of such 
states is $N$. These multi-particle states are dynamically localized due to the
kicking, and this remains true even when interactions are taken into account. 
The flat dispersion for these states is shown in Fig.~\ref{fig:effnondisp} for 
some representative cases; we find that the eigenvalues of the Floquet 
operator obtained numerically agree very well with the analytical expression.

\begin{figure}[H]
\centering
\includegraphics[width=14cm]{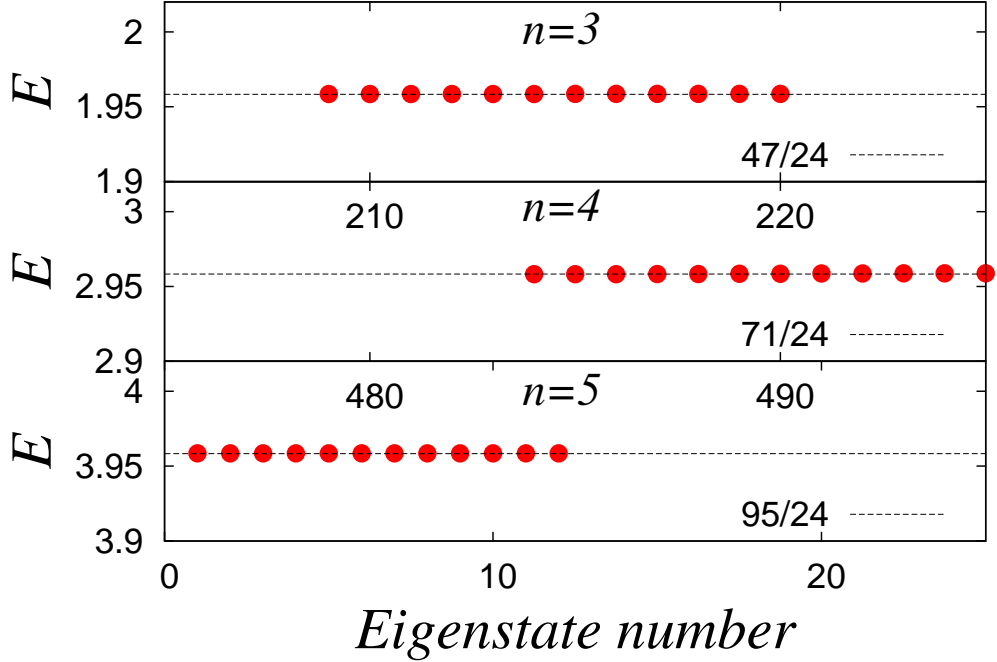}
\caption{Numerically obtained eigenvalues of the Floquet operator as compared 
with the analytical expression in Eq.~\eqref{en}, for $V=1, ~T= 0.5$ and
$\ga =1$, for $n=3,
~4, ~5$ particles on $12$ sites. Note that for each $n$ we have $N=12$ 
eigenvalues which are non-dispersing.} \mylabel{fig:effnondisp} \end{figure}

\begin{figure}[H]
\centering
\includegraphics[width=14cm]{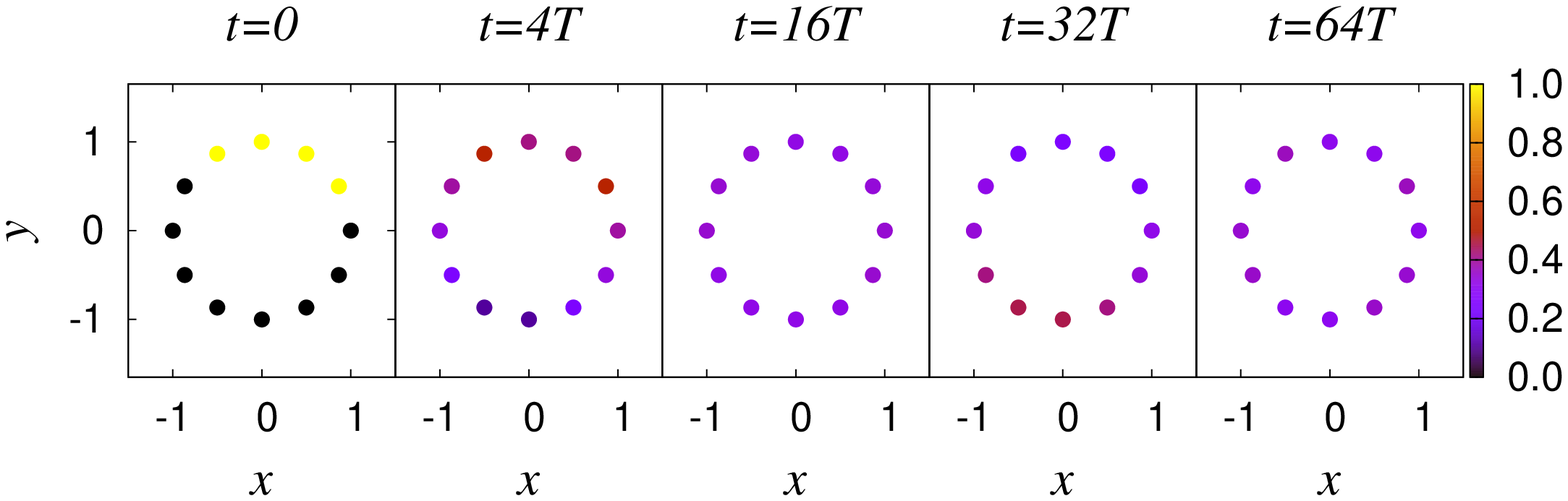}
\includegraphics[width=14cm]{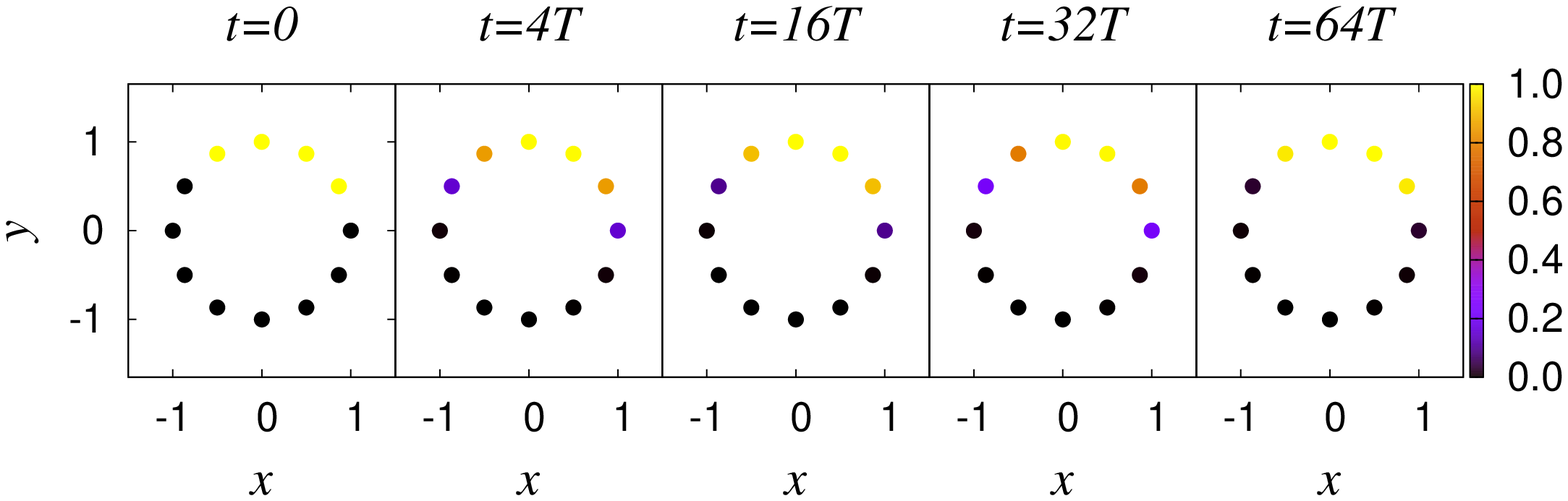}
\caption{Time evolution of a four-particle state: (i) $V=1$, no 
kicking, (ii) $V=1$, with kicking. In both cases $\ga = 1$ and $T=0.5$.
There are four particles on 12 sites, and they are initially located on four 
adjacent sites. The second row shows that the particles do not move even in 
the presence of interactions.} \label{fig:dyn5} \end{figure}

As a striking demonstration of the dynamical localization of multi-particle
systems, we show the time evolution of a system with four particles on
12 sites in Fig.~\ref{fig:dyn5}. We see that the particles remain dynamically
localized when they are initially located on four adjacent sites.

\section{Spin-$1/2$ fermions with on-site interactions}
\label{sec_spinhalf}

We now look at the one-dimensional model of spin-1/2 interactions
with on-site interactions between spin-up and spin-down electrons.
This is called the Hubbard model and it is also exactly solvable by
the Bethe ansatz~\cite{mattis93,sutherland04}. The Hamiltonian of the model is
\beq H ~= - \ga ~\sum_{j,\si} (c_{j\si }^\dg c_{j+1~\si} + H. c.) ~+~
U ~\sum_j ~n_{j\ua} n_{j\da}. \label{ham2} \eeq
We naturally identify the first term as $H_{NI}$ and the second term as $H_I$.
As before, we first evaluate $[H_{NI},H_I]$ which has relevant terms, 
\bea &=& [c_{j\si}^\dg c_{j+1 ~\si} + c^\dg_{j+1~\si}c_{j\si}, 
n_{j \ua} n_{j \da} + n_{j+1 ~\ua} n_{j+1 ~\da} ] \non \\
&=& (c^\dg_{j+1 ~\ua}c_{j\ua} -c_{j\ua}^\dg c_{j+1 ~\ua}) (n_{j\da} - 
n_{j+1~\da}) + (c^\dg_{j+1 ~\da}c_{j\da} -c_{j\da}^\dg c_{j+1 ~\da}) (n_{j\ua} 
- n_{j+1~\ua}). \non \\
\eea
Next we find 
\bea && [H_{NI},[H_{NI},H_I]] ~=~ \ga^2 T^2 U \non \\
&\times& \Big[ 4 ~\Big((n_{j\ua} - n_{j+1 ~\ua})(n_{j \da}- n_{j+1~\da}) + 
(c^\dg_{j\ua}c_{j+1 ~\ua}- c^\dg_{j+1~\ua}c_{j\ua})(c^\dg_{j\da}
c_{j+1 ~\da} - c^\dg_{j+1~\da}c_{j\da}) \Big) \non \\
&& ~+ \Big( \big((c^\dg_{j\ua}c_{j+2 ~\ua} + c^\dg_{j+2~\ua} c_{j\ua})
(n_{j\da}-n_{j+1~\da}) \non \\
&& ~~~~~~~+ (c^\dg_{j\ua} c_{j+1 ~\ua}-c^\dg_{j+1~\ua}c_{j\ua})
(c^\dg_{j+2~\da} c_{j+1~\da} - c^\dg_{j+1~\da}c_{j+2~\da}) \big) 
+ \big(\ua \leftrightarrow \da \big) \Big) \non \\
&& ~+ \Big( \big((c^\dg_{j+1~\ua}c_{j-1 ~\ua} + c^\dg_{j-1~\ua} c_{j+1~\ua})
(-n_{j\da}+n_{j+1~\da}) \non \\
&& ~~~~~~~+ (c^\dg_{j\ua} c_{j+1 ~\ua}-c^\dg_{j+1~\ua}c_{j\ua}) (c^\dg_{j\da} 
c_{j-1~\da} - c^\dg_{j-1~\da}c_{j\da}) \big) + \big(\ua \leftrightarrow 
\da\big) \Big) \Big]. \eea

The effective Hamiltonian in \eqref{EffH} therefore takes the form 
\bea H_{eff} &=& U ~\sum_j ~n_{j\ua} n_{j\da} \non \\ 
&&- \frac{i\ga TU}{2} \sum_{j, \si} (c^\dg_{j+1 ~\si}c_{j\si} -c_{j\si}^\dg 
c_{j+1 ~\si}) (n_{j\bar{\si}} - n_{j+1~\bar{\si}}) \non \\
&&- \frac{\ga^2 T^2 U}{6} \sum_j \Big[ 4 \big((n_{j\ua} - n_{j+1 ~\ua})
(n_{j\da}-n_{j+1~\da}) \non \\
&&~~~~~~~~~~~~~~~~~+ (c^\dg_{j\ua}c_{j+1 ~\ua}- c^\dg_{j+1~\ua}c_{j\ua}) 
(c^\dg_{j\da} c_{j+1~\da} - c^\dg_{j+1~\da}c_{j\da}) \big) \non \\
&&~~~~~~~~~~~~~~~~~+ \left((c^\dg_{j\ua}c_{j+2 ~\ua} + c^\dg_{j+2~\ua} 
c_{j\ua}) (n_{j\da} -n_{j+1~\da}) + (\ua \leftrightarrow \da)\right) \non \\
&&~~~~~~~~~~~~~~~~~- \left((c^\dg_{j-1~\ua}c_{j+1 ~\ua} + c^\dg_{j+1~\ua} 
c_{j-1~\ua}) (n_{j\da}-n_{j+1~\da}) + (\ua \leftrightarrow \da)\right) \non \\
&&~~~~~~~~~~~~~~~~~+ \Big(2(c^\dg_{j\ua}c_{j+1 ~\ua}-c^\dg_{j+1~\ua}
c_{j\ua}) (c^\dg_{j+2~\da} c_{j+1~\da} - c^\dg_{j+1~\da} c_{j+2~\da}) +
(\ua \leftrightarrow \da) \Big) \Big]. \non \\
\label{eq:effHspin} \eea

We now use the effective Hamiltonian in Eq.~\eqref{eq:effHspin} to look at
two-particle states. In particular, we will again search for bound states. 
In the Hubbard model, two particles can interact with each other only if 
they have opposite spins. We will therefore take the two particles to have 
spins $\ua$ and $\da$.

We first look at a state where the two particles are at the same site $j$.
(This is a spin singlet state). A momentum eigenstate will be of the form
\beq |\psi_k \rangle ~=~ \sum_j ~e^{ikj} | j \ua, j\da \rangle. \eeq
(For $k=\pi$, this is again an exact eigenstate of both the Hamiltonian 
in Eq.~\eqref{ham2} and of the kicking problem since $N_A = 0$ or 2 implies 
that $U_K |\psi_k \rangle = |\psi_k \rangle$).
We will look at the effect of each of the terms in Eq.~\eqref{eq:effHspin}
on the state $| j \ua, j\da \rangle$. This is shown in Table IV, with a
sum over $j$ being assumed.

\begin{table}[H]
\begin{center}
\begin{tabular}{|c|c|}
\hline 
Terms in $H_{eff}$ & Acting on $e^{ikj}|j\ua,j\da\rangle$ \\ 
\hline 
$U n_{j\ua} n_{j\da}$ & $U e^{ikj}|j\ua,j\da \rangle$ \\ 
\hline 
$-\frac{i\ga T U}{2}(c^\dg_{j+1 ~\si}c_{j\si} -c_{j\si}^\dg c_{j+1 ~\si}) 
(n_{j\bar{\si}} - n_{j+1~\bar{\si}})$ & $- \frac{i\ga TU}{2} e^{ikj} 
(|j\ua, j+1 ~\da \rangle + |j\ua,j-1 ~\da \rangle$ \\ 
& ~~~~~~ $+ |j+1~\ua,j\da \rangle + |j-1~\ua,j \da\rangle)$ \\ 
\hline
$- \frac{\ga^2 T^2 U}{6}\Big[ 4 \big((n_{j\ua} - n_{j+1 ~\ua}) (n_{j\da}-
n_{j+1~\da})$ & $- \frac{2 \ga^2 T^2 U}{3} e^{ikj} ( 2|j\ua, j\da \rangle$ \\ 
$+ (c^\dg_{j\ua} c_{j+1 ~\ua}-c^\dg_{j+1~\ua}c_{j\ua})(c^\dg_{j\da} 
c_{j+1~\da} - c^\dg_{j+1~\da}c_{j\da}) \big)$ & $+ |j+1~\ua, j+1~\da \rangle
+ |j-1~\ua, j-1~\da \rangle)$ \\ 
\hline
$+ \left((c^\dg_{j\ua}c_{j+2 ~\ua} + c^\dg_{j+2~\ua} c_{j\ua}) (n_{j\da}-
n_{j+1~\da}) + (\ua \leftrightarrow \da)\right)$ & $-$ \\ 
$- \left((c^\dg_{j-1~\ua}c_{j+1 ~\ua} + c^\dg_{j+1~\ua} c_{j-1~\ua}) (n_{j\da}
-n_{j+1~\da}) + (\ua \leftrightarrow \da)\right)$ & \\ 
$+ \Big(2(c^\dg_{j\ua}c_{j+1 ~\ua}-c^\dg_{j+1~\ua} c_{j\ua}) (c^\dg_{j+2~\da}
c_{j+1~\da} - c^\dg_{j+1 ~\da}c_{j+2~\da})$ & \\
$+ (\ua \leftrightarrow \da) \Big) \Big]$ & \\
\hline
\end{tabular}
\end{center}
\caption{Effect of various terms in $H_{eff}$ on the state $e^{ikj} |j \ua, 
j \da \rangle$. The $-$ symbol in the right column means we have states which 
only contribute to the bound state at orders higher than $\ga^2 T^2 U$.} 
\end{table}

{} From Table IV, we see that the terms of order $\ga TU$ can give rise
to a second order process where an initial state $|j \ua, j \da \rangle$
can go to intermediate states $|j \ua, j \pm 1 ~\da \rangle$ and then
return to $|j \ua, j \da \rangle$. The contribution of this is
\beq \frac{\ga^2 T^2 U^2}{4} ~2e^{ikj} (1+e^{ik})(1+e^{-ik}) ~| j\ua, j \da
\rangle \eeq
divided by the energy difference between the initial and intermediate states
which is $U$. We therefore get 
\beq \ga^2 T^2 U (1+\cos k). \eeq
To this we add the contribution of the terms of order $\ga^2 T^2 U$ which
is equal to
\beq - \frac{4\ga^2 T^2 U}{3} (1+\cos k). \eeq
The total quasienergy is therefore
\bea E_{1k} &=& U ~+~ \ga^2 T^2 U ~(1-\frac{4}{3}) ~(1+\cos k) \non \\
&=& U ~-~ \frac{2 \ga^2 T^2 U}{3} ~\cos^2 \left(\frac{k}{2} \right). 
\label{e1k2} \eea
This is the quasienergy for a wave function in which there is a large amplitude
for the particles with up and down spins to be at the same site. 

We now look at a different case where the two particles with opposite spins 
(to be denoted as $\si$ and $\bar \si$) are at adjacent sites $j$ and $j+1$.
The wave function with momentum $k$ is then
\beq | \psi_k \rangle ~=~ \sum_{j\si} e^{ik(j+1/2)} s_\si |j\si,j+1~\bar{\si} 
\rangle, \label{psik1} \eeq
where $s_\si = +1$ if $\si = \ua$ and $-1$ if $\si = \da$. (This is again
a spin singlet state). The action of the different terms in 
Eq.~\eqref{eq:effHspin} on the wave function in Eq.~\eqref{psik1} is shown 
in Table V, with a sum over $j$ and $\si$ being assumed.

\begin{table}[H]
\begin{center}
\begin{tabular}{|c|c|}
\hline 
Terms in $H_{eff}$ & Acting on $e^{ik(j+1/2)} s_\si |j\si,j+1~\bar{\si} 
\rangle$ \\ 
\hline 
$U n_{j\ua} n_{j\da}$ & $zero$ \\ \hline $-\frac{i\ga T U}{2}(c^\dg_{j+1 ~\si}
c_{j\si} -c_{j\si}^\dg c_{j+1 ~\si}) (n_{j\bar{\si}} - n_{j+1~\bar{\si}})$ & 
$i\ga T U e^{ik(j+1/2)} (|j\ua, j \da \rangle$ \\
& $+ |j+1~\ua,j+1~\da \rangle)$ \\
\hline
$- \frac{\ga^2 T^2 U}{6}\Big[ 4 \big((n_{j\ua} - n_{j+1 ~\ua}) (n_{j\da}-
n_{j+1~\da})$ & $\frac{4 \ga^2 T^2 U}{3} e^{ik(j+1/2)} s_\si |j \si, j+1 ~
\bar{\si} \rangle$ \\
$+ (c^\dg_{j\ua} c_{j+1 ~\ua}-c^\dg_{j+1~\ua}c_{j\ua})(c^\dg_{j\da} 
c_{j+1~\da} - c^\dg_{j+1~\da}c_{j\da}) \big)$ & \\ 
\hline
$+ \left((c^\dg_{j\ua}c_{j+2 ~\ua} + c^\dg_{j+2~\ua} c_{j\ua})(n_{j\da}-
n_{j+1~\da}) + (\ua \leftrightarrow \da) \right)$ 
& $\frac{2\ga^2 T^2 U}{3} e^{ik(j+1/2)} s_\si ~\times$ \\ 
$- \left( (c^\dg_{j-1~\ua}c_{j+1 ~\ua} + c^\dg_{j+1~\ua} c_{j-1~\ua})
(n_{j\da}-n_{j+1~\da}) + (\ua \leftrightarrow \da) \right)$ & $(|j+1 ~\si, 
j+2 ~\bar{\si} \rangle +|j-1 ~\si, j \bar{\si} \rangle)$ \\ 
$+ \Big(2(c^\dg_{j\ua}c_{j+1 ~\ua}-c^\dg_{j+1~\ua} c_{j\ua})(c^\dg_{j+2~\da}
c_{j+1~\da} - c^\dg_{j+1 ~\da}c_{j+2~\da})$ & \\ $+(\ua \leftrightarrow \da) 
\Big) \Big]$ & \\ 
\hline
\end{tabular}
\end{center}
\caption{Effect of various terms in $H_{eff}$ on the state $e^{ik(j+1/2)}
s_\si |j\si,j+1~\bar{\si} \rangle$.}
\end{table}

Table V shows that the term of order $\ga TU$ takes an initial state $s_\si 
|j \si, j+1 \bar{\si} \rangle$ to an intermediate state $|j \si, j \bar{\si}
\rangle$ and then back to the initial state. This gives a contribution equal to
\beq \frac{\ga^2 T^2 U^2}{4} ~2e^{ik(j+1/2)} (1+e^{ik})(1+e^{-ik}) s_\si
| j\si, j+1~ \bar{\si} \rangle. \eeq
Dividing by a denominator $-U$ equal to the energy difference of the two 
states, we get 
\beq - \ga^2 T^2 U ~(1+\cos k). \eeq
Adding the contribution from the terms of order $\ga^2 T^2 U$, we get a total
contribution equal to 
\bea E_{2k} &=& (\frac{4}{3}-1)\ga^2 T^2 U ~(1+\cos k) \non \\
&=& \frac{2\ga^2 T^2 U}{3}~\cos^2 \left(\frac{k}{2} \right). \label{e2k2} \eea

In Fig.~\ref{fig:eff2} we compare the numerically obtained eigenvalues of 
the Floquet operator for two particles with spins $\ua$ and $\da$ with the 
analytical expressions in Eqs.~\eqref{e1k2} and \eqref{e2k2}. The agreement 
can be seen to be excellent.

\begin{figure}[H]
\centering
\includegraphics[width=14cm]{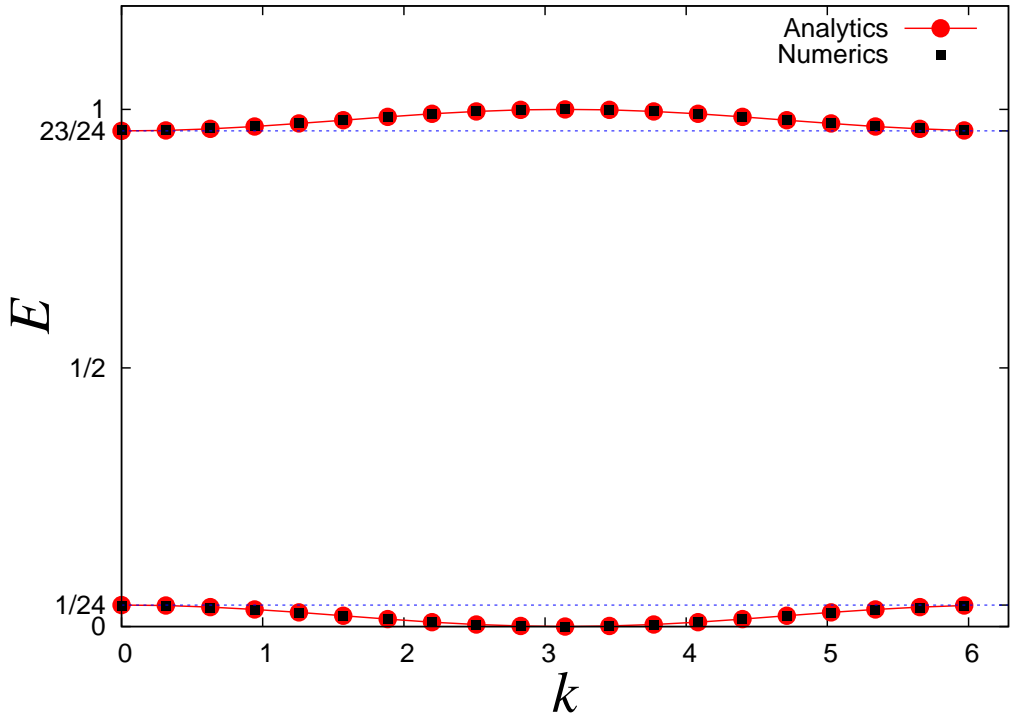}
\caption{Numerically obtained eigenvalues of the effective Hamiltonian as 
compared with the analytical expressions in Eqs.~\eqref{e1k2} and \eqref{e2k2}, 
for $U=1, ~T= 0.25$ and $\ga =1$. All other eigenvalues are zero. We have one 
$\ua$ and one $\da$ particle on $20$ sites.} 
\mylabel{fig:eff2} \end{figure}

\begin{figure}[H]
\centering
\includegraphics[width=14cm]{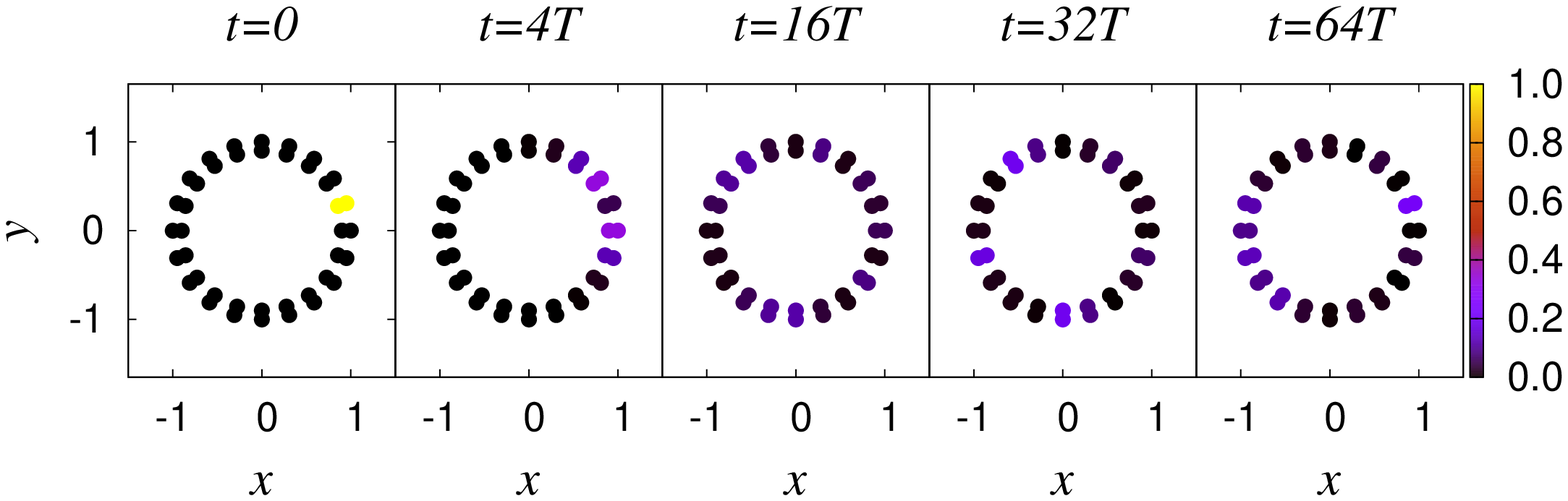}
\includegraphics[width=14cm]{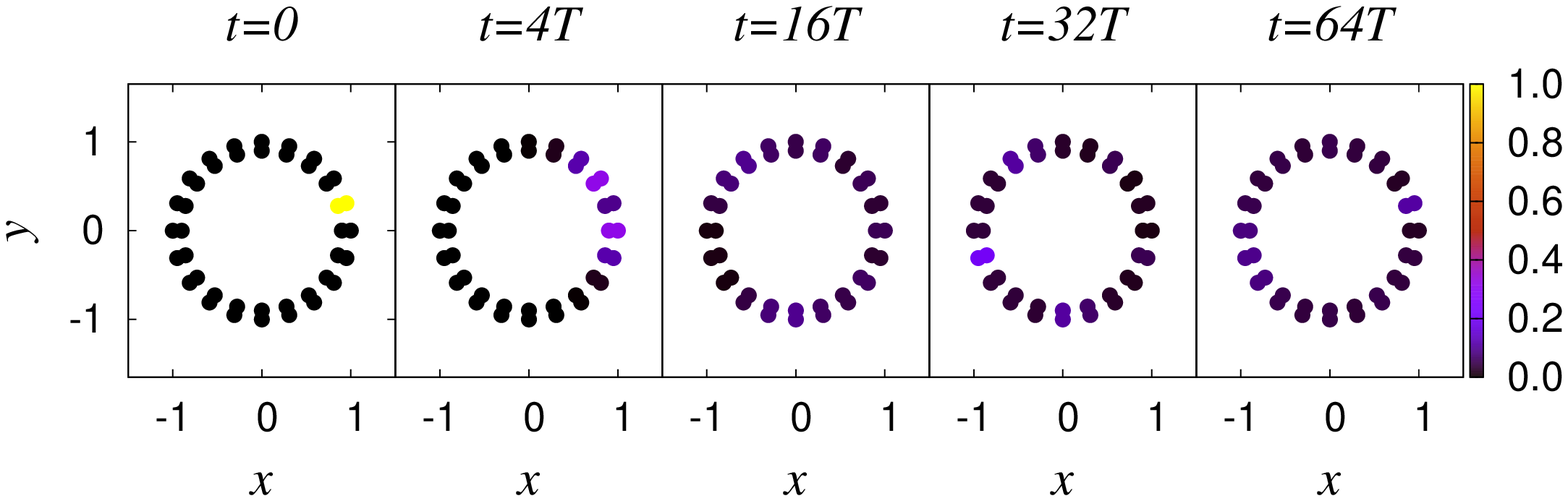}
\includegraphics[width=14cm]{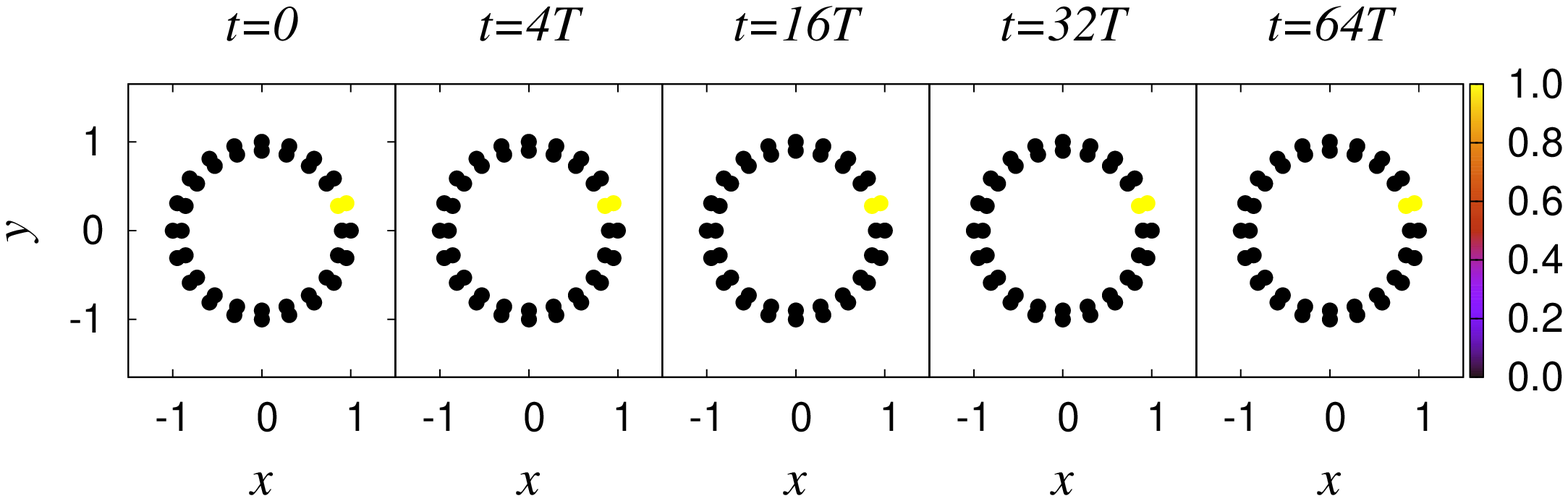}
\includegraphics[width=14cm]{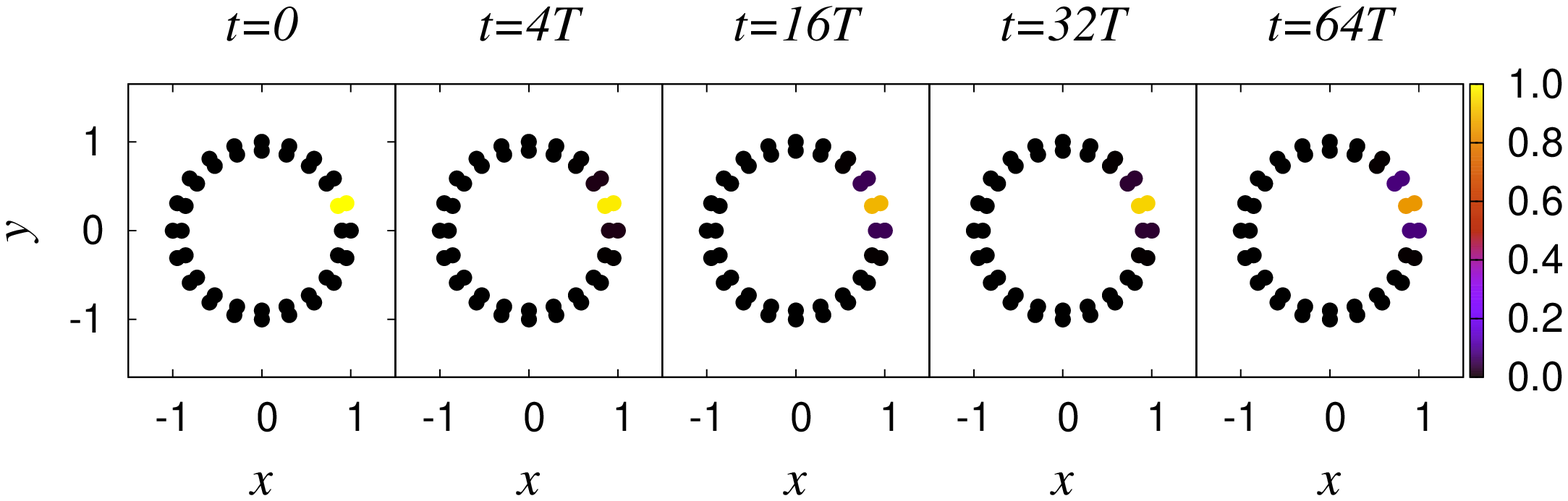}
\caption{Time evolution of a two-particle state for 
four cases: (i) $U=0$, no kicking, (ii) $U=1$, no kicking, (iii) $U=0$, with 
kicking, (iv) $U=1$, with kicking. In all cases $\ga = 1$ and $T=0.25$.
There are two particles, with spins $\ua$ and $\da$, on 20 sites, and they 
are initially located at the same site. The colors of the dots on the outer 
(inner) ring show the expectation values of the number of up (down) spin 
particles at different sites.} \label{fig:dyn6} \end{figure}

In Fig.~\ref{fig:dyn6}, we show the time evolution of a system with two
particles, with spins $\ua$ and $\da$, on 20 sites; the particles are 
initially at the same site. The third row shows that the particles are 
dynamically localized when there is kicking but no interactions. The
fourth row shows that when interactions are turned on, the particles move 
but very slowly; this is because the group velocity for the dispersion 
in Eq.~\eqref{e1k2} is small when $\ga^2 T^2$ is small.

\section{Bosons with on-site interactions}
\label{sec_bose}

As our final example of an interacting system, we will consider a system
of bosons with on-site interactions in one dimension. This is called the
Bose Hubbard model. For a system with $N$ sites and periodic boundary 
conditions, the Hamiltonian is
\beq H ~=~ \sum_{j=1}^N ~[-\ga (b_j^\dg b_{j+1} + H. c.) ~+~ \frac{U}{2} 
n_j (n_j-1)], \label{hbos} \eeq
where $n_j = b_j^\dg b_j$ is the particle number at site $j$.

As before we first evaluate 
\bea [H_{NI},H_I] &=&
-\ga U \left((b^\dg_{j+1} n_j b_j - b^\dg_j n_j b_{j+1}) - (b^\dg_{j+1} n_{j+1} 
b_j - b^\dg_j n_{j+1} b_{j+1} ) \right) \non \\
&=& -\ga U \left(b^\dg_{j+1} (n_j - n_{j+1})b_j - b^\dg_j (n_j - n_{j+1}) 
b_{j+1} \right). \eea
The next term is 
\bea [H_{NI},[H_{NI},H_I]] &=& \ga^2 T^2 U \Big[ 2 \Big( n_j (n_j-1) + 
n_{j+1}(n_{j+1}-1) - 4 n_j n_{j+1} \Big) \non \\
&& + 2 \Big(b^\dg_j b_{j+1}b^\dg_j 
b_{j+1} + H. c. \Big) \non \\
&& + \Big(b^\dg_j (n_j - 2n_{j+1}) b_{j+2} + H. c. \Big) + \Big(b^\dg_j b_{j+1}
b^\dg_{j+2} b_{j+1} + H. c.\Big) \non \\
&& + \Big( b^\dg_{j-1}(n_{j+1} - 2n_j) b_{j+1} + H. c. \Big) + 
\Big(b^\dg_{j-1} b_j b^\dg_{j+1} b_j + H. c.\Big) \Big]. \eea

Putting all this together, the effective Hamiltonian in Eq.~\eqref{EffH}
takes the form
\bea H_{eff} &=& \frac{U}{2} ~\sum_j ~n_j(n_j-1) \mylabel{efbo1} \\ 
&-& \frac{i\ga T U}{2} ~\sum_j ~\left( b^\dg_{j+1} (n_j - n_{j+1})b_j - b^\dg_j
( n_j - n_{j+1}) b_{j+1} \right) \mylabel{efbo2} \\
&-& \frac{\ga^2 T^2 U}{3} ~\sum_j ~\Big[ 2 n_j(n_j-1) - 4 n_j n_{j+1} 
\mylabel{efbo3} \\
&+& \Big( b^\dg_j b_{j+1}(b^\dg_j + b^\dg_{j+2} )b_{j+1} + H. c. \Big) ~+~ 
\frac{1}{2} \Big( b^\dg_j(n_{j+2} + n_j - 4n_{j+1}) b_{j+2} + H. c. \Big) 
\Big]. \mylabel{efbo4} \eea

We can again look for two-particle bound states just as in the previous 
sections. We first look for a state with momentum $k$ which consists 
mainly of states in which both the particles are at site $j$, namely,
\beq | \psi_{1k} \rangle ~=~ \sum_j ~e^{ikj} |j,j\rangle. \label{jj} \eeq
(For $k=\pi$, this is an exact eigenstate of the Hamiltonian in 
Eq.~\eqref{hbos}) and of the kicking problem since $U_K | \psi_{1k} \rangle = 
| \psi_{1k} \rangle$. The action of $H_{eff}$ on the state in \eqref{jj} is 
given in Table VI.

\begin{table}[H]
\begin{center}
\begin{tabular}{|c|c|}
\hline 
Terms in $H_{eff}$ & Acting on $e^{ikj}|j,j\rangle$ \\ 
\hline 
$\frac{U}{2} ~n_j(n_j-1)$ & $Ue^{ikj}|j,j\rangle$ \\ 
\hline 
$- \frac{i\ga T U}{2} ~\left( b^\dg_{j+1} (n_j - n_{j+1})b_j - b^\dg_j
( n_j - n_{j+1}) b_{j+1} \right)$ & $- \frac{i\ga T U}{2} \sqrt{2} e^{ikj}
\left(|j,j+1\rangle + |j-1,j\rangle\right)$ \\ 
\hline
$- \frac{\ga^2 T^2 U}{3} ~\Big[ 2 n_j(n_j-1) - 4 n_j n_{j+1}$ & 
$- \frac{4\ga^2 T^2 U}{3}e^{ikj}|j,j\rangle$ \\ 
\hline
$+ \Big( b^\dg_j b_{j+1}(b^\dg_j + b^\dg_{j+2} )b_{j+1} + H. c. \Big) 
$ & $- \frac{2\ga^2 T^2 U}{3}e^{ikj}(|j-1,j-1\rangle +|j+1,j+1\rangle)$ \\ 
\hline
$+ ~\frac{1}{2} \Big( b^\dg_j(n_{j+2} + n_j - 4n_{j+1}) b_{j+2} + H. c. \Big) 
\Big]$ & $-$ \\ 
\hline
\end{tabular}
\end{center}
\caption{Effect of various terms in $H_{eff}$ on the state $e^{ikj}
|j,j\rangle$.}
\end{table}

The terms in the second line in Table VI take $|j,j\rangle$ to an 
intermediate state $|j,j\pm 1\rangle$ and act again to 
take it back to $|j,j\rangle$ with a contribution 
\beq \frac{\ga^2 T^2 U^2}{4} ~2e^{ikj} (1+e^{ik})(1+e^{-ik}) | j, j \rangle.
\eeq
Dividing by the energy difference between the initial and intermediate states, 
$U$, gives the contribution 
\beq \ga^2 T^2 U (1+\cos k). \eeq
The third and fourth lines in Table VI give a diagonal contribution of the
form
\beq - ~\frac{4\ga^2 T^2 U}{3} ~(1+\cos k). \eeq
The total contribution to the quasienergy is therefore
\bea E_{1k} &=& U ~+~ \ga^2 T^2 U ~(1-\frac{4}{3}) ~(1+\cos k) \non \\
&=& U ~-~ \frac{2 \ga^2 T^2 U}{3} ~\cos^2 \left(\frac{k}{2} \right). 
\label{e1k3} \eea

We now look at the second kind of two-particle bound states which consists
mainly of states where the particles are on sites $j$ and $j+1$, namely,
\beq |\psi_{2k} \rangle ~=~ \sum_j ~e^{ik(j+1/2)} ~|j,j+1 \rangle. \eeq
The action of $H_{eff}$ on this state is given in Table VII.

\begin{table}[H]
\begin{center}
\begin{tabular}{|c|c|}
\hline 
Terms in $H_{eff}$ & Acting on $e^{ik(j+1/2)}|j,j+1\rangle$ \\ 
\hline 
$\frac{U}{2} ~n_j(n_j-1)$ & $zero$ \\ \hline 
$- \frac{i\ga T U}{2} ~\left( b^\dg_{j+1} (n_j - n_{j+1})b_j - b^\dg_j (n_j 
- n_{j+1}) b_{j+1} \right)$ & $\frac{i\ga T U}{2} \sqrt{2} e^{ik(j+1/2)}
\left(|j,j\rangle + |j+1,j+1\rangle\right)$ \\ 
\hline
$- \frac{\ga^2 T^2 U}{3} ~\Big[ 2 n_j(n_j-1) - 4 n_j n_{j+1}$ & $\frac{4\ga^2
T^2 U}{3}e^{ik(j+1/2)}|j,j+1\rangle$ \\ 
\hline
$+ \Big( b^\dg_j b_{j+1}(b^\dg_j + b^\dg_{j+2} )b_{j+1} + H. c. \Big)
$ & $-$ \\ 
\hline
$+~ \frac{1}{2} \Big( b^\dg_j(n_{j+2} + n_j - 4n_{j+1}) b_{j+2} + H. c. \Big) 
\Big]$ & $\frac{2\ga^2 T^2 U}{3}e^{ik(j+1/2)}(|j-1,j\rangle +|j+1,j+2\rangle)$
\\ 
\hline
\end{tabular}
\end{center}
\caption{Effects of various terms in $H_{eff}$ on the state $e^{ik(j+1/2)} 
|j,j+1 \rangle$.}
\end{table}

The second line in Table VII takes $|j,j+1\rangle$ to intermediate states
$|j,j\rangle$ and $|j+1,j+1 \rangle$, and acts again to take it back to 
$|j,j+1\rangle$ with a contribution
\beq \frac{\ga^2 T^2 U^2}{4} ~2e^{ikj} (1+e^{ik})(1+e^{-ik}) | j, j+1 \rangle.
\eeq
Dividing by the energy difference between the initial and intermediate states, 
$-U$, gives the contribution
\beq -\ga^2 T^2 U (1+\cos k). \eeq
To this we have to add the contributions from the third and fifth lines of 
Table VII. The total quasienergy is therefore
\bea E_{2k} &=& \ga^2 T^2 U ~(\frac{4}{3}-1) ~(1+\cos k) \non \\
&=& \frac{2 \ga^2 T^2 U}{3} ~\cos^2 \left(\frac{k}{2} \right). 
\label{e2k3} \eea

We note that the dispersions given in Eqs.~\eqref{e1k3} and \eqref{e2k3}
are identical to Eqs.~\eqref{e1k2} and \eqref{e2k2}. A comparison between 
the numerically obtained eigenvalues of the effective Hamiltonian 
and the analytical expressions in Eqs.~\eqref{e1k3} and \eqref{e2k3}
therefore looks exactly the same as in Fig.~\ref{fig:eff2} if we take the
same values of $U, ~T$ and $\ga$.


Finally, we find that just as in the case of spinless fermions, we have 
$n$-particle bound states which are dynamically localized and which do not 
disperse if $n \ge 3$; such bound states consist mainly of states in
which all the $n$ particles are on the same site $j$. For a system of
$N$ sites, there are $N$ such bound states corresponding to the different
possible values of $j$. The quasienergy of these states is given by
\bea E_n ~=~ \frac{U}{2} n (n-1) ~\left( 1 ~-~ \frac{\ga^2T^2}{3} \right). \eea
We have verified that our numerical results for $n$-particle states
match this analytical expression.

\subsection{Effective Hamiltonian when each site has a double degeneracy}

We will now consider what happens if a uniform potential is applied at all 
sites (this is equivalent to applying a chemical potential $\mu$) in such a 
way that, in the absence of periodic driving, the ground 
state of the interaction part of the Hamiltonian has a 
two-fold degeneracy at each site corresponding to occupancies $p$ and $p+1$; 
here $p$ can be $0, 1, 2, \cdots$. (These are the points where the Mott 
lobes meet in the phase diagram of the Bose Hubbard model in the limit
of zero hopping~\cite{freericks94}). Namely, we modify the 
interaction term in Eq.~\eqref{hbos} to 
\beq \frac{U}{2}(n_j -c)^2, ~~{\rm where}~~ c ~=~ p ~+~ \frac{1}{2}, \eeq
so that the states with $n_j = p$ and $p+1$ are degenerate with energy $U/8$.
We then find that the effective Hamiltonian is given by 
Eqs.~(\ref{efbo1}-\ref{efbo4}) except that Eq.~\ref{efbo1} is now replaced by 
$\frac{U}{2} ~\sum_j ~(n_j- p - \frac{1}{2})^2$.


We will now assume $U$ is so large that the energies of the states with 
$n_j = p$ and $p+1$ are well separated from the energies of states 
with any other value of $n_j$. With this assumption, we will turn on the
periodic driving and derive an effective Hamiltonian $H_{eff}$ in the 
space of states in which $n_j =p$ or $p+1$ at each site. To this end,
we introduce pseudo-spin Pauli matrices $\si^a_j$ at each site 
(where $a = x,y,z$), so that the states with $n_j = p$ and $p+1$ correspond
to $\si^z_j = -1$ and $+1$ respectively. Hence
\beq n_j ~=~ p ~+~ \frac{1+ \si^z_j}{2}. \label{nj} \eeq
Further, within the space of these two states, we have the identities 
\bea b_j^\dg ~=~ \sqrt{p+1} ~\si_j^+ ~~~~{\rm and}~~~~ b_j ~=~ 
\sqrt{p+1} ~\si_j^-. \eea

We will derive $H_{eff}$ up to order $\ga^2 T^2 U$. As before there are 
two kinds of contributions: those coming from second order processes induced 
by the terms of order $\ga T U$ in Eq.~\eqref{efbo2}, and those coming directly
from the terms of order $\ga^2 T^2 U$ in Eqs.~(\ref{efbo3}-\ref{efbo4}).
The second order processes can lead to terms in $H_{eff}$ which involve
either two sites or three sites. We present the details of the calculation in 
Appendix \ref{Bosondetails}. The effective Hamiltonian is found to be
\bea H_{eff} 
&=& \frac{\ga^2 T^2 U}{12} ~\sum_j ~\Big[ 2 (p+1) \si^z_j ~+~ \si^z_j 
\si^z_{j+1} ~+~ (p+1)(p+1+\si^z_{j+1})(\si^+_j\si^-_{j+2} + H. c.) \non \\
&& ~~~~~~~~~~~~~~~~~~+ ~(2p^2+4p+1) \Big]. \label{heffp} \eea

\subsection{Highly degenerate eigenstates for the case $p=0$}

We now consider the special case $p=0$ for the effective Hamiltonian in
Eq.~\eqref{heffp}, namely, the states with $n_j = 0$ and $1$ are degenerate 
for the interaction part of the Hamiltonian in \eqref{hbos}. We then get
\bea H_{eff} &=& \frac{\ga^2 T^2 U}{12} ~\sum_j ~\left[ (1+\si_j^z)(1+
\si_{j+1}^z) ~+~ (1+ \si^z_{j+1})(\si^+_j\si^-_{j+2} + \si^-_j \si^+_{j+2}) 
\right]. \label{eq:effhbos} \eea
It turns out that this has an exponentially large number of degenerate
eigenstates with zero quasienergy. This can be shown as follows. 

We first consider a local Hamiltonian defined as
\beq H_j ~=~ (1+ \si^z_j) \left[ \frac{1}{2} (1+ \si^z_{j-1}) ~+~ \frac{1}{2}
(1+ \si^z_{j+1}) ~+~ \si_{j-1}^+ \si^-_{j+1} ~+~ \si_{j-1}^- \si^+_{j+1} 
\right]. \label{hj} \eeq
It is easy to find the eigenvalues of $H_j$ since it only involves 
three spins and therefore eight states. We find that the eigenstates have a
six-fold degeneracy with eigenvalue zero and a two-fold 
degeneracy with eigenvalue 4. Further, all the states in which two 
neighboring sites (either $j-1, ~j$ or $j, ~j+1$) do {\it not} both have 
$\si^z_n = +1$ are eigenstates with zero eigenvalue.

Next, we note that the Hamiltonian in \eqref{eq:effhbos} can be written as 
a sum of the Hamiltonians in \eqref{hj},
\beq H_{eff} ~=~ \frac{\ga^2 T^2 U}{12} ~\sum_j ~H_j. \eeq
Given this structure, it can be shown that if there is a state which is an 
eigenstate of each of the $H_j$'s simultaneously, then it is also an 
eigenstate of $H_{eff}$; further, the eigenvalue of $H_{eff}$ is equal to 
the sum of the eigenvalues of all the $H_j$'s. (The
opposite is not necessarily true; an eigenstate of $H_{eff}$ need not be
an eigenstate of each of the $H_j$'s). It follows from this and the statement
made above about the eigenstates of $H_j$ that {\it any} state in which no 
two neighboring sites have $\si^z_n = +1$ is an eigenstate state of $H_{eff}$,
and the corresponding eigenvalue (quasienergy) is zero. 

If the number of sites $N$ is large, one can use the transfer matrix 
method~\cite{pathria96} 
to find the number of states in which two sites with $\si_n^z = +1$ are
not next to each other. Consider the one-dimensional Ising model in a
magnetic field whose strength is such that the Hamiltonian takes the form
\beq H_{Ising} ~=~ J ~\sum_j ~(1 + \si_j^z) ~(1 + \si_{j+1}^z), 
\label{ising} \eeq
where $J > 0$. The four possible states for two neighboring sites $j$ and 
$j+1$ have the energies $4J$ when both sites have $\si_n^z = +1$ and 
zero for the other three cases. Hence the eigenstates of Eq.~\eqref{ising}
also have the property that two neighboring sites must not both have $\si_n^z 
= +1$. The partition function of this system at an inverse temperature $\beta$
is given by
\beq Z (\beta) ~=~ tr \left[ \left( \begin{array}{cc}
e^{-4\beta J} & 1 \\
1 & 1 \end{array} \right)^N \right] \eeq
for a periodic system with $N$ sites. In the limit $\beta \to \infty$, the 
partition function gives the number of eigenstates. For large $N$, we 
see that the number of eigenstates grows exponentially as
\beq Z (\infty) ~=~ tr \left[ \left( \begin{array}{cc}
0 & 1 \\
1 & 1 \end{array} \right)^N \right] ~\simeq~ \tau^N, \eeq
where $\tau = (\sqrt{5}+1)/2$ is the golden ratio. This is a lower bound on
the eigenstate degeneracy since there may be other eigenstates of 
$H_{eff}$ which are not of the form described above.

Before ending this section, we note that our analysis of the large
number of degenerate 
eigenstates that we have found for the effective Hamiltonian derived up to 
order $\ga^2 T^2 U$ is only valid up to some finite time scale; beyond that 
time, higher order effects will become important and the system may
eventually heat up~\cite{bilitewski15,genske15,kuwahara16}.

\section{Effects of perturbations on dynamical localization}
\label{sec_pert}

In this section, we will consider various perturbations and study how far the 
phenomenon of dynamical localization is robust against them. We will ignore 
the effects of interactions in this section. Hence the discussion below will 
be the same for bosons and fermions.

We consider non-interacting spinless particles in one dimension with
nearest-neighbor hopping. This is a bipartite system with the Hamiltonian
\beq H ~=~ ~-~ \ga ~\sum_{n=1}^N ~[c_n^\dg c_{n+1} ~+~ H.c.], \label{ham3} \eeq
where we have assumed that the system has $N$ sites (we will take $N$ to
be even), and we use periodic boundary 
conditions. We Fourier transform to momentum space as
\bea c_k &=& \frac{1}{\sqrt N} ~\sum_{n=1}^N ~e^{-ikn} ~c_n, \non \\
c_n &=& \frac{1}{\sqrt N} ~\sum_{-\pi < k \le \pi} ~e^{ikn} ~c_k, \eea
where $k$ goes from $-\pi$ to $+\pi$ in steps of $2\pi/N$. Then 
Eq.~\eqref{ham3} can be written as
\beq H ~=~ \sum_{-\pi < k \le \pi} ~(-2 \ga ~\cos k) ~c_k^\dg c_k. 
\label{ham4} \eeq

As one example of a perturbation, we consider what happens if this
system is kicked by an operator of the form in Eq.~\eqref{uk},
\beq U_K ~=~ e^{-i \al N_A}, \label{uk2} \eeq
where $\al$ can be different from $\pi$. If we take the $A$ sublattice
to be the sites corresponding to even values of $n$, we have
\bea N_A &=& \sum_{even ~n} ~c_n^\dg c_n ~=~ \sum_{all ~n} ~
\frac{1}{2} ~(1 ~+~ (-1)^n) ~c_n^\dg c_n \non \\
&=& \sum_{-\pi < k \le \pi} ~\frac{1}{2} ~(c_k^\dg c_k ~+~ 
c_{k+\pi}^\dg c_k). \label{na} \eea
In the two-level space given by $k$ and $k+\pi$, we can write Eqs.~\eqref{ham4}
and \eqref{na} as
\bea H &=& \sum_{0 \le k < \pi} ~\left( \begin{array}{cc}
c_k^\dg & c_{k+\pi}^\dg 
\end{array} \right) ~(-2 \ga \cos k) ~\si^z ~\left( \begin{array}{c}
c_k \\
c_{k + \pi} \end{array} \right), \non \\
N_A &=& \sum_{0 \le k < \pi} ~\left( \begin{array}{cc}
c_k^\dg & c_{k+\pi}^\dg 
\end{array} \right) ~\frac{1}{2} ~(I ~+~ \si^x) ~\left( \begin{array}{c}
c_k \\
c_{k + \pi} \end{array} \right), \label{hna} \eea
respectively, where $I, ~\si^x$ and $\si^z$ denote identity and Pauli matrices
in pseudo-spin space. Since the pair of modes $(k,k+\pi)$ (where $0 \le k <
\pi$) corresponding to different values of $k$ are decoupled from each other, 
we can consider the different values of $k$ separately. Following 
Eq.~\eqref{hna} we define two matrices
\bea h_k ~=~ (-2 \ga \cos k) ~\si^z ~~~~{\rm and}~~~~ n_{ak} ~=~ \frac{1}{2} ~
(I ~+~ \si^x). \eea
The Floquet operator for one time period for momentum $k$ is then given by
\beq U_k ~=~ \exp [- ~\frac{i \al}{2} ~(I ~+~ \si^x)]~ \exp [ i 2 \ga T 
\cos k~ \si^z]. \label{uk3} \eeq

Writing the eigenvalues of $U_k$ in Eq.~\eqref{uk3} as $e^{\pm i \ep_k T}$,
where $\ep_k$ is the quasienergy, we find that 
\beq \ep_k ~=~ -~ \frac{1}{T} ~\cos^{-1} [\cos (\frac{\al}{2}) 
\cos (2 \ga T \cos k)] ~+~ \frac{\al}{2T}. \eeq
For $\al = 0$ (no kicking), we recover the usual dispersion $\ep_k = - 2 \ga
\cos k$ with group velocity given by $v_g = |d\ep_k /dk| = 2 \ga \sin k$, 
while for $\al = \pi$ (dynamical localization), we obtain $\ep_k = 0$
with group velocity $v_g = 0$ for all $k$. In general we have
\beq v_g (k) ~=~ \frac{2 \ga \cos (\frac{\al}{2}) ~|\sin (2 \ga T \cos k) 
\sin k|}{\sqrt{1 ~-~ \cos^2 (\frac{\al}{2}) \cos^2 (2 \ga T \cos k)}}. 
\label{vgk} \eeq
For some given values of $\al$ and $\ga T$, it is convenient to define a 
quantity $v_{max}$ as the maximum value of $v_g$ in the range $0 \le k \le 
\pi$. This has the physical meaning of being the maximum velocity (called the 
Lieb-Robinson bound) with which information can propagate in the 
system~\cite{lieb72}. We will see below that $v_{max}$ plays an important 
role. For $\al$ close to $\pi$, we can see from Eq.~\eqref{vgk} that 
$v_{max}$ is of order $|\pi - \al|$.

In Fig.~\ref{fig:vmax} the solid red line shows a plot of $v_{max}$ 
versus $\al$ for $T = 0.5$ and $\ga =1$ as obtained from Eq.~\eqref{vgk}; we 
see that $v_{max}$ smoothly goes from 2 to zero as $\al$ goes from zero to 
$\pi$. The black squares in Fig.~\ref{fig:vmax} show the maximum velocity 
derived from a numerical study of the propagation of a particle at long times
as discussed below.

\begin{figure}[H]
\centering
\includegraphics[width=14cm]{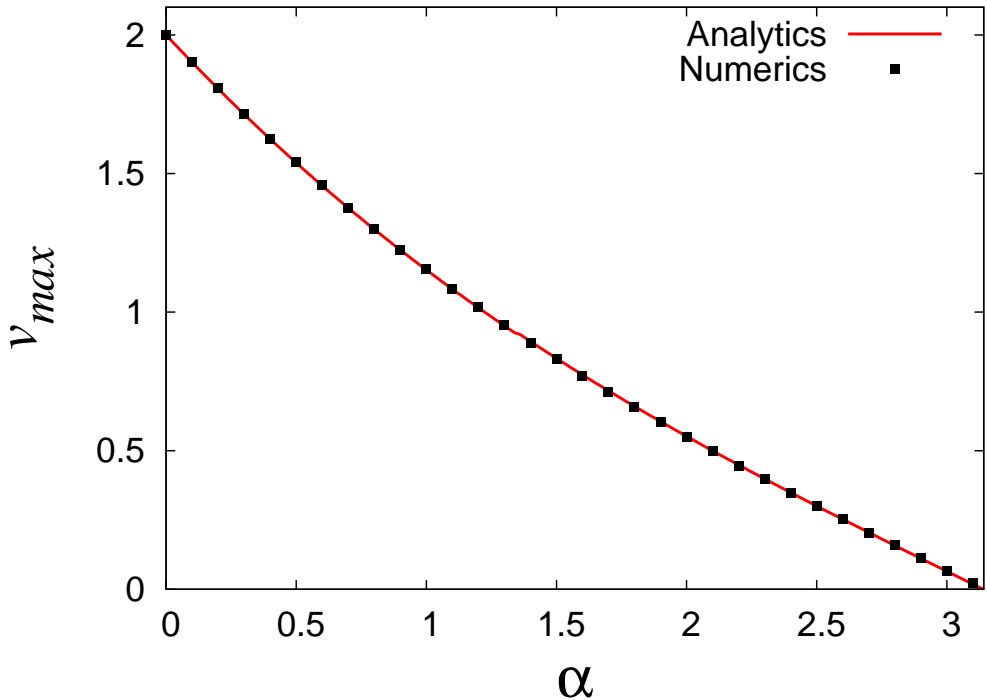}
\caption{Plot of $v_{max}$ versus $\al$ for $T = 0.5$ and $\ga = 1$. The 
solid red line shows the analytical result obtained from Eq.~\eqref{vgk}, while
the black squares show the result obtained numerically from a study of the
propagation of a particle as discussed in the text.} 
\mylabel{fig:vmax} \end{figure}

We now study the time evolution of a one-particle state, where the particle 
is initially at one particular site in the middle of a long chain with 
$N$ sites. Taking this site to be $n=0$, the initial state is given by 
\beq | \psi (0) \rangle ~=~ \int_{-\pi}^\pi ~\frac{dk}{2\pi} ~| k \rangle, 
\label{psi0} \eeq
where we have taken the limit $N \to \infty$ so that $k$ is now a continuous
variable. Upon evolving this for a time $T$ (but before acting with a 
$\de$-function kick), $|k\rangle \to e^{i 2 \ga T \cos k} |k \rangle$.
The wave function at site $n$ is then
\beq \psi_n (T) ~=~ \int_{-\pi}^\pi ~\frac{dk}{2\pi} ~e^{i(kn + 2 \ga T 
\cos k)}. \label{psiT} \eeq
This integral gives a Bessel function~\cite{abram72} and we find that
the probability of finding the particle at site $n$ is
\beq |\psi_n (T)|^2 ~=~ |J_{|n|} (2 \ga T) |^2. \label{jn} \eeq
This probability remains unchanged when the particle is then given a kick
with an arbitrary strength $\al$, since a kick only changes the phase of
$\psi_n$ by $e^{-i\al}$ on sites belonging to the $A$ sublattice.
We therefore conclude that Floquet evolution for one time period spreads out 
the probability from the initial value of 1 at site $n=0$ to the expression 
given in Eq.~\eqref{jn}.

For a given value of $2 \ga T$, it is known that $J_{|n|} (2 \ga T)$ rapidly 
goes to zero when $|n|$ becomes much larger than $2 \ga T$. 
Namely,~\cite{abram72}
\beq J_{|n|} (2 \ga T) ~\sim~ \frac{1}{\sqrt{2 \pi |n|}} ~\left( \frac{e 
\ga T}{|n|} \right)^{|n|} \eeq
for $|n| \gg 2 \ga T$. Eq.~\eqref{jn} therefore implies that the particle 
spreads out a distance of the order of $2 \ga T$ in time $T$; this is 
consistent with the fact that $v_{max} = 2 \ga$ for a particle with the 
dispersion $\ep_k = 2 \ga \cos k$. To make this more precise, we calculate the 
square of the width of the wave function at time $t$,
\beq m_2 (t) ~\equiv~ \sum_{n=-\infty}^\infty ~n^2 ~|\psi_n (t)|^2. \eeq
Using the identity $\sum_{n=1}^\infty n^2 [J_n (x)]^2 = x^2/4$ for real
$x$, we see from Eq.~\eqref{jn} that
\beq m_2 (T) ~=~ \frac{1}{2} ~v_{max}^2 T^2, \label{m2} \eeq
where $v_{max} = 2 \ga$.

We now study what happens to $m_2$ at integer multiples of $T$ up to very 
large times. Fig.~\ref{fig:m2} shows a plot of $m_2$ versus $t=nT$ for 
$\al = 3.12$, $T = 0.5$ and $\ga = 1$. Since $\al$ is close to $\pi$, the
particle should be almost dynamically localized. We indeed see that $m_2$
remains of order 1 up to a large time $t$ although there are pronounced
oscillations between odd and even integer values of $t/T$. Beyond that large
time, however, odd and even integer values of $t/T$ give the same values of
$m_2$. For such large times, a fit of the form 
\beq m_2 ~=~ A ~t^p \label{fit} \eeq
gives $p=2.0$. Fig.~\ref{fig:vmax} compares the dependence of $v_{max}$ on 
$\al$ as obtained analytically from Eq.~\eqref{vgk} (solid red line) and the 
dependence of $\sqrt{2A}$ on $\al$ as found numerically by fitting the large
time behavior in Fig.~\ref{fig:m2} to the form in Eq.~\eqref{fit} (black 
squares), for $\ga =1$ and $T=0.5$. The fact that the two match perfectly 
means that the parameter $A$ in Eq.~\eqref{fit} is equal to $v_{max}^2/2$ 
for all values of $\al$.

We can understand the time-dependence of $m_2$ for both small and large 
times as follows. We begin with Eq.~\eqref{uk3}. For $\al$ close to $\pi$, 
the leading order form of $U_k$ is given by
\bea U_k &\simeq& \exp [- ~\frac{i \pi}{2} ~(I ~+~ \si^x)]~ \exp 
[ i 2 \ga T \cos k~ \si^z] \non \\
&=& - ~\cos (2 \ga T \cos k) ~\si^x ~-~ \sin (2 \ga T \cos k) ~\si^y. 
\label{uk4} \eea
Acting with $U_k$ on the column $(1,1)^T$ (which corresponds to the initial
wave function $|k\rangle + |k +\pi \rangle$ given in Eq.~\eqref{psi0}), 
we get $(-e^{-i2 \ga T \cos k}, -e^{i2 \ga T \cos k})^T$ which corresponds
to the wave function $-e^{-i2 \ga T \cos k} |k\rangle -e^{i2 \ga T \cos k} 
|k +\pi \rangle$. This is the same as the wave function in Eq.~\eqref{psiT};
this implies that $m_2 (T) = (2 \ga T)^2 /2$. Next, Eq.~\eqref{uk4} 
implies that $U_k^2 \simeq I$. We therefore have $U_k^{2p+1} \simeq U_k$ 
while $U_k^{2p} ~\simeq I$ for any integer $p$. This would imply that $\psi 
((2p+1)T) \simeq \psi (T)$ so that $m_2 ((2p+1)T) \simeq (2 \ga T)^2 /2$, while
$\psi (2pT) \simeq \psi (0)$ so that $m_2 (2pT) \simeq 0$. Thus $m_2$ is
expected to alternate between $(2 \ga T)^2 /2$ and a small number as $t/T$
alternates between odd and even integers. This agrees with what we see
in Fig.~\ref{fig:m2} till $t/T$ reaches a large value of about 90; beyond
this time $m_2$ has the same value for odd and even integer values of $t/T$
and increases quadratically with $t$. We can estimate the value of $t/T$
where this behavior begins as follows.

For $\al = \pi - \eta$, where $\eta$ is small, we find from Eq.~\eqref{uk3} 
that 
\beq U_k^2 ~=~ e^{i\eta} ~\exp [i \eta \cos (2 \ga T \cos k) ~\{ \cos (2 \ga T
\cos k) ~\si^x ~+~ \sin (2 \ga T \cos k) ~\si^y \}] \label{uk5} \eeq
up to first order in $\eta$. We can compare this with the value of $U_k$ for 
$\al = \pi$ which, from Eq.~\eqref{uk4}, is given by
\beq U_k ~=~ i~ \exp [ \frac{i\pi}{2} ~\{ \cos (2 \ga T \cos k) ~\si^x ~+~ 
\sin (2 \ga T \cos k) ~\si^y \}]. \label{uk6} \eeq
We have seen above, time evolution with $U_k$ gives $m_2 (T) = (2 \ga T)^2 /2$.
Ignoring the $k$-independent phases in Eqs.~(\ref{uk5}-\ref{uk6}) which do not 
affect the value of $m_2$, we see that the form of $U_k^{2p}$ will become 
identical to the form of $U_k$ when $2p = t/T$ is given by
\beq p \eta \cos (2 \ga T \cos k) ~=~ \frac{\pi}{2}. \eeq
It is clear that the value of $p$ depends on $k$. However, the ballistic
motion that is visible for $t \gtrsim 90$ in Fig.~\ref{fig:m2} is dominated
by the values of $k$ where $v_g (k) = v_{max}$. For $\al = 3.12$ (hence
$\eta = \pi - 3.12$), $T=0.5$ 
and $\ga=1$, we find from Eq.~\eqref{vgk} that $v_g = v_{max}$ for $k=0.829$ 
and $2.313$ (these add up to $\pi$). At these values of $k$, we have 
$\cos (2 \ga T \cos k) = 0.780$; we then get $p = (\pi/2)/(0.780 \times \eta) 
= 93$. We see from Fig.~\ref{fig:m2} that $t \simeq 2 \times 93 \times T = 93$
does approximately give the point at which the values of $m_2$ for odd and 
even integer values of $t/T$ merge and the ballistic motion begins.

\begin{figure}[H]
\centering
\includegraphics[width=14cm]{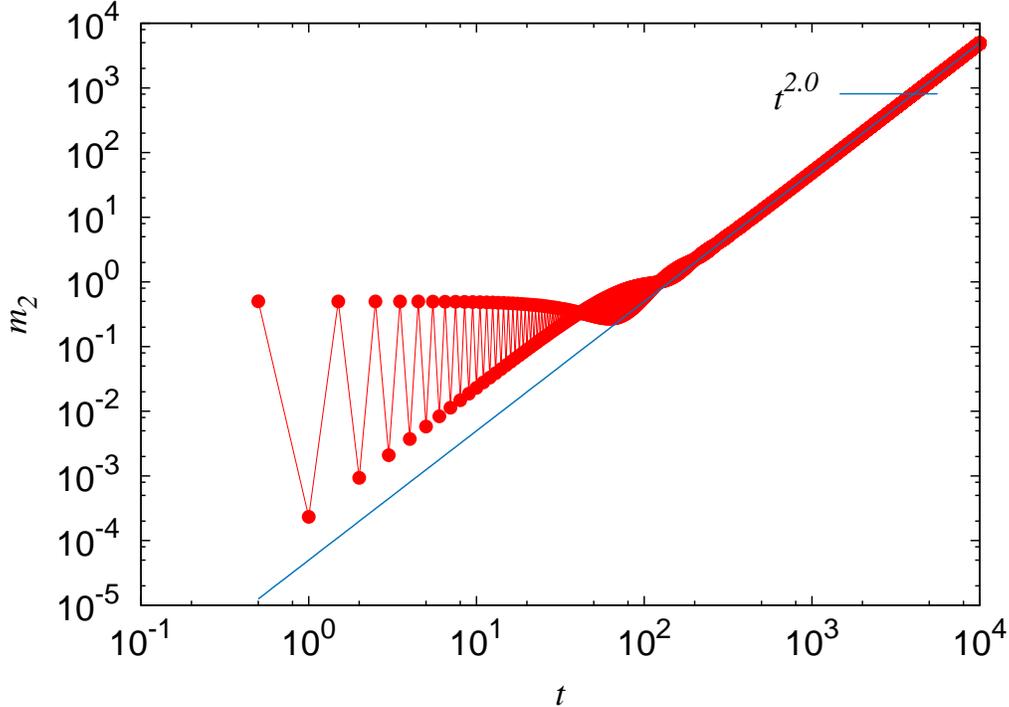}
\caption{Plot of $m_2$ versus $t$ for $\al = 3.12$, $T = 0.5$ and $\ga = 1$. 
The particle is initially at a site in the middle of a system with 2000 sites.
At short times, $m_2$ alternates between two values depending on whether $t/T$
is an odd or even integer. At long times, a power law fit between $m_2$ and 
$t$ shows that $m_2$ increases as $t^{2.0}$, implying that the particle is 
moving ballistically.} \mylabel{fig:m2} \end{figure}

We conclude that for $\al$ close to $\pi$, a single particle remains
dynamically localized up to a large time of order $1/|\pi - \al|$; up to this
time $m_2$ alternates between two values, one of order $(2 \ga T)^2/2$ and
the other of order zero, for odd and even values of $t/T$. Beyond that
large time, $m_2$ increases quadratically with time indicating that the
particle moves ballistically with a velocity $v_{max}$ which is of order
$|\pi - \al|$. (The initial oscillations in $m_2$ are similar to those seen
for other quantities in some recent papers on Floquet time 
crystals~\cite{lazarides14d,keyser16,else16a,else16b}).

As another example of a perturbation, we consider what happens if there
is disorder in the hopping amplitudes and the system
is given $\de$-function kicks with $\al = \pi$. Namely, the Hamiltonian is 
\beq H ~=~ \sum_{n=1}^N ~t_{n,n+1} ~(c_n^\dg c_{n+1} ~+~ H.c.), \eeq
where $t_{n,n+1}$ can have some randomness.
If this is kicked with an operator of the form $U_K = e^{-i \pi N_A}$, we find
that the time evolution operator for two time periods is given by
\bea U^2 &=& e^{-i \pi N_A} ~e^{- i HT} ~e^{-i \pi N_A} ~e^{- i HT} \non \\
&=& I \label{uk7} \eea
since $e^{-i \pi N_A}$ anticommutes with $(c_n^\dg c_{n+1} + H.c.)$.
Hence a particle will be dynamically localized after every integer multiple 
of $2T$.

\section{Concluding remarks}
\label{sec_concl}

In this paper, we have examined the effects of interactions in bipartite 
lattice systems where periodic $\de$-function kicks applied to the sublattice 
potential with a strength $\al = \pi$
lead to dynamical localization if we view the system stroboscopically. We 
have shown that interactions can generate new kinds of hoppings between 
nearest- and next-nearest-neighbor sites which depend on the occupation 
numbers on some nearby sites. These hoppings give rise to a variety of 
interesting effects.

We began by describing a formalism for calculating the effective Floquet 
Hamiltonian in an expansion in powers of $T$. We then calculated the
Hamiltonian to second order in $T$ in three different models in one dimension.
For spinless fermions with a nearest-neighbor interaction $V$, we showed that
the two-particle sector has two branches of bound states: one 
branch which has a dispersion lying around $V$ and another 
branch with a dispersion around zero. We further showed that there are 
$n$-body bound states, with $n \ge 3$, which are dispersionless; hence they 
do not move with time.
For the Hubbard model of spin-1/2 fermions with an on-site interaction $U$, 
we showed that the two-particle spin singlet sector has two branches of
bound states with dispersions lying close to $U$ and zero respectively.
In this model we do not find any $n$-body bound states if $n \ge 3$.
For the Bose Hubbard model of bosons with on-site interaction $U$, we
again found two branches of two-particle bounds states with quasienergies close
to $U$ and zero, and dispersionless $n$-body bound states if $n \ge 3$. 
We also studied a special case of this model in which 
the interactions make states with occupancies $p$ and $p+1$ degenerate at 
each site. This allowed us to define a pseudo-spin-1/2 degree of freedom
at each site, and we found an effective Hamiltonian which lies in the
subspace of these states. For $p=0$, we obtained a particularly simple
form of the effective Hamiltonian. We showed that a class of eigenstates
of the effective Hamiltonian can be found exactly, and the degeneracy of 
the corresponding quasienergy grows exponentially with the system size.
Finally, we showed that if the kicking strength $\al$ is slightly
different from $\pi$, a particle remains dynamically localized for a long
time of the order of $1/|\pi - \al|$ but then moves ballistically with a
maximum velocity of the order of $|\pi - \al|$.

Turning to possible experimental realizations of the models studied in this 
paper, we note that a dynamical localization-to-delocalization transition has 
been observed in a quantum kicked rotor. Such a system is realized by placing
cold atoms in a pulsed standing wave; the transition is detected by measuring 
the number of atoms which have zero velocity when a quasiperiodic driving
is applied~\cite{ringot00}. Given that cold atom systems provide a versatile
platform for simulating a wide variety of condensed matter systems, our
paper shows that a combination of periodic driving and interactions can 
lead to a variety of remarkable phenomena.

We would like to end by pointing out some possible directions for future
studies.

\noi (i) It would be interesting to study if dynamical localization induced 
by periodic driving along with interactions can give rise to topological 
phases. We note that in Ref.~\onlinecite{mikami16}, it was shown that 
circularly polarized light (which corresponds to simple harmonic driving) can 
give rise to transitions to topological phases; the effect of interactions was
then studied within dynamical mean-field theory. One can similarly investigate
if periodic $\de$-function kicks and interactions can drive topological phase
transitions.

\noi (ii) A generalization of our results to bipartite lattice models
in higher dimensions may be interesting. It is not difficult to carry out 
a perturbative expansion of the effective Hamiltonian in any dimension. 
However, it may be more difficult to find bound states of two or more 
particles and to study the time evolution of few-particle states
in higher than one dimension. 

\noi (iii) We have mainly concentrated on the dynamics of 
systems with a small number of particles. (An exception to this was 
the analysis in Secs. \ref{sec_bose} A and B where we looked at 
systems with an arbitrary number of particles). It may be useful 
to study the thermodynamics of a system with a finite filling fraction
of particles. In particular, one can look at the possible 
phases of such systems (for instance, if they are metals, superfluids or 
insulators) and the nature of the excitations in the different 
phases. We note that such a study requires us to couple the system to
a thermal reservoir, and the phases of the system may depend on the form 
of the system-reservoir couplings~\cite{iadecola15a,iadecola15b}. Some
recent papers have studied scattering processes and heating effects in
periodically driven systems with interactions~\cite{bilitewski15,genske15}.

\noi (iv) We have seen in some cases that there are few-particle bound 
states with a dispersionless spectrum. This raises the question of whether
the spectrum would continue to be so simple if we expand the effective 
Hamiltonian to higher than second order in $T$. Another interesting
question to ask is: what is the time scale up to which the results 
obtained from the effective Hamiltonian derived to order $T^2$ remain
accurate? An answer to this has been provided in Ref.~\onlinecite{kuwahara16}
where a time scale is found up to which the results obtained using an
effective Hamiltonian derived to order $T^n$ and the exact Floquet operator 
match well and beyond which they start disagreeing.

We also know that the models of interacting spinless and spin-1/2 fermions 
are Bethe ansatz solvable~\cite{mattis93,sutherland04}. We may investigate if 
this has any implications for the properties of the system in the presence 
of periodic $\de$-function kicking.


\noi (v) It is interesting to compare our results with those found in 
many-body localization (MBL). In MBL, the localization is due to the spatial 
disorder and/or interactions. Some 
studies have then looked at the effects of periodic driving on the MBL 
state~\cite{ponte15a,lazarides14c,ponte15b}. Our motivation and study
are completely distinct. We begin with a system which is completely 
dynamically localized even in the absence of disorder. We then probe the
effect of interactions on systems with a few particles. 
The few-particle bound states that we find are again
dynamically localized. Unlike MBL systems, the driving protocol plays the
essential role here of localizing the particles. It would be interesting
to study an interplay of dynamical localization due to driving, spatial
localization due to disorder, and interactions.


\section*{Acknowledgments}

A.A. thanks Sambuddha Sanyal for some discussions. We thank H. Katsura, 
T. Kuwahara and A. Lazarides for useful comments. A.A. thanks Council of 
Scientific and Industrial Research, India for funding through a SRF 
fellowship. D.S. thanks Department of Science and Technology, India for 
Project No. SR/S2/JCB-44/2010 for financial support.

\appendix

\section{Mathematical Identities}

We begin with the identity
\beq e^X e^Y ~=~ e^{Y + [X,Y] + \frac{1}{2!}[X,[X,Y]] + \frac{1}{3!}[X,[X,
[X,Y]]] + \cdots} ~e^X. \eeq
If $[X,Y] = \ga Y$, where $\ga$ is a number, then the above equation implies
that
\beq e^X e^Y = e^{(e^\ga) Y}e^X. \mylabel{eq:expiden} \eeq
If $[X,Z]=0$ along with $[X,Y] = \ga Y$, then we get
\beq e^X e^{Y+Z} ~=~ e^{(e^\ga) Y + Z} ~e^X. \eeq

The Baker-Campbell-Hausdorff formula gives
\beq e^X e^Y ~=~ e^{X+Y+\frac{1}{2}[X,Y] + \frac{1}{12}([X,[X,Y]]+[Y,[Y,X]]) + 
\cdots}, \eeq
which implies that
\beq \ln (e^X e^Y) ~=~ X ~+~ Y ~+~ \frac{1}{2}[X,Y] ~+~ \frac{1}{12}([X,[X,Y]]
~+~[Y,[Y,X]]) ~+~ \cdots. \mylabel{eq:BCH} \eeq

If $X=C+D$ and $Y=C-D$, then
\bea \ln (e^{C+D} e^{C-D}) &=& 2C ~+~ [D,C] ~+~ \frac{1}{3}([C,D]D+D[D,C]) ~+~ 
\cdots \non \\
&=& 2C ~+~ [D,C] ~+~ \frac{1}{3}[D,[D,C]] ~+~ \cdots. \mylabel{eq:BCHSD} \eea

Finally, for fermion operators we know that
\beq [ n_j, c_j ] ~=~ -c_j ~~~{\rm and}~~~ [ n_j, c^\dg_j ] = c^\dg_j,
\mylabel{eq:iden} \eeq
where $n_j = c^\dg_j c_j$. For bosons
\beq [b_i,b^{\dg}_j] ~=~ \de_{ij}, \eeq
and this gives the same commutation relations between $n_j = b^\dg_j b_j$
and $b_j, ~b^\dg_j$ as in Eq.~\eqref{eq:iden}.

\section{Derivation of effective Hamiltonian for the bosonic model}
\label{Bosondetails}

We now present the details of the calculation of the effective 
Hamiltonian when the occupancies $p$ and $p+1$ of a site are degenerate.

\noi {\bf Second order processes involving two sites:}
\vspace*{.1cm}

The various processes will be shown below as tables. Each table will show
an initial (or intermediate) state $I$ and an intermediate (or final) state 
$F$, with $I_j$ and $F_j$ denoting the number of particles at site $j$ in 
the $I$ and $F$ states respectively.
\vspace*{.2cm}

\noi 1. 
\vspace*{-1cm}

\beq
\begin{array}{|c|c||c|c||c|}
\hline
I_j & I_{j+1} & F_j & F_{j+1} & Amplitude \\
\hline 
\hline
p & p & p-1 & p+1 & \frac{i\ga T U}{2} ~\sqrt{p}\sqrt{p+1} \\
p-1 & p+1 & p & p & -\frac{i\ga T U}{2} ~\sqrt{p}\sqrt{p+1} \\
\hline
\end{array}
\eeq
\vspace*{.1cm}

\bitem
\item The energy denominator coming from the difference of the unperturbed
energies of the initial and final states is $-U$.
\item This process can occur in two ways, as we can have $F_j = p+1, ~F_{j+1}
= p-1$. So we get a total contribution
\beq 2 ~\left( \frac{i\ga T U}{2} \sqrt{p}\sqrt{p+1}\right) ~\left( 
\frac{-i\ga T U}{2} \sqrt{p}\sqrt{p+1} \right) ~\left( \frac{1}{-U}\right) ~
=~ -\frac{p(p+1)\ga^2 T^2 U}{2}. \eeq
\eitem
\vspace*{.2cm}

\noi 2. 
\vspace*{-1cm}

\beq
\begin{array}{|c|c||c|c||c|}
\hline
I_j & I_{j+1} & F_j & F_{j+1} & Amplitude \\
\hline 
\hline
p & p+1 & p-1 & p+2 & i \ga T U ~\sqrt{p}\sqrt{p+2} \\
p-1 & p+2 & p & p+1 & -i \ga T U ~\sqrt{p}\sqrt{p+2} \\
\hline
\end{array}
\eeq
\vspace*{.1cm}

\bitem
\item The energy denominator is $-2U$. 
\item The total contribution is
\beq -\frac{p(p+2) \ga^2 T^2 U}{2}. \eeq
\item A similar process occurs when the initial state has $I_j = p+1, ~I_{j+1} 
= p$.
\eitem
\vspace*{.2cm}

\noi 3. 
\vspace*{-1cm}

\beq
\begin{array}{|c|c||c|c||c|}
\hline
I_j & I_{j+1} & F_j & F_{j+1} & Amplitude \\
\hline 
\hline
p+1 & p+1 & p & p+2 & \frac{i\ga TU}{2} ~\sqrt{p+1}\sqrt{p+2} \\
p & p+2 & p+1 & p+1 & -\frac{i\ga TU}{2} ~\sqrt{p+1}\sqrt{p+2} \\
\hline
\end{array}
\eeq
\vspace*{.1cm}

\bitem
\item The energy denominator is $-U$.
\item This process can occur in two possible ways. So the total contribution is
\beq -\frac{(p+1)(p+2) \ga^2 T^2 U}{2}. \eeq
\eitem
\vspace*{.2cm}

We now find that all the above terms can be fitted to an expression of the form
\beq a_1 ~\si^z_j ~+~ a_2~\si^z_{j+1} ~+~ a_3 ~\si^z_j \si^z_{j+1} ~+~ a_4. 
\eeq
Comparing this expression with the contributions given above, we obtain
\bea -a_1 - a_2 + a_3 + a_4 &=& - \frac{\ga^2 T^2 U}{2} ~p(p+1), \non \\
-a_1 + a_2 - a_3 + a_4 &=& - \frac{\ga^2 T^2 U}{2} ~p(p+2), \non \\
a_1 - a_2 - a_3 + a_4 &=& - \frac{\ga^2 T^2 U}{2} ~p(p+2), \non \\
a_1 + a_2 + a_3 + a_4 &=& - \frac{\ga^2 T^2 U}{2} ~(p+1)(p+2). \eea
These imply
\bea a_1 &=& -\frac{\ga^2 T^2 U}{4} ~(p+1), \non \\
a_2 &=& -\frac{\ga^2 T^2 U}{4} ~(p+1), \non \\
a_3 &=& -\frac{\ga^2 T^2 U}{4}, \non \\
a_4 &=& -\frac{\ga^2 T^2 U}{4} ~(2p^2+4p+1). \eea
 
We therefore have the following terms in $H_{eff}$ so far 
\beq - \frac{\ga^2 T^2 U}{4} ~\Bigl[ ~(p+1) ~(\si^z_j + \si^z_{j+1}) ~+~ 
\si^z_j \si^z_{j+1} ~+~ (2p^2+4p+1) \Bigr]. \label{hlow1} \eeq
\vspace*{.1cm}

\noi {\bf Second order processes involving three sites:}
\vspace*{.2cm}

\noi 1.
\vspace*{-1cm}

\beq
\begin{array}{|c|c|c||c|c|c||c|}
\hline
I_j & I_{j+1} & I_{j+2} & F_j & F_{j+1} & F_{j+2} & Amplitude \\
\hline
p & p & p & - & - & - & - \\
\hline
\end{array}
\eeq
\vspace*{.2cm}

The symbol $-$ in the table means that the terms in Eq.~\eqref{efbo2}
take the state $(I_j,I_{j+1},I_{j+2})$ to a state which is not relevant 
to the calculation of $H_{eff}$.
\vspace*{.2cm}

\noi 2. 
\vspace*{-1cm}

\beq
\begin{array}{|c|c|c||c|c|c||c|}
\hline
I_j & I_{j+1} & I_{j+2} & F_j & F_{j+1} & F_{j+2} & Amplitude \\
\hline 
\hline
p & p & p+1 & p+1 & p-1 & p+1 & \frac{i\ga TU}{2} ~\sqrt{p}\sqrt{p+1} \\
p+1 & p-1 & p+1 & p+1 & p & p & - \frac{i\ga TU}{2} ~\sqrt{p}\sqrt{p+1} \\
\hline
\end{array}
\eeq
\vspace*{.2cm}

\bitem
\item The energy denominator is $-U$.
\item The total contribution is 
\beq -\frac{p(p+1) \ga^2 T^2 U}{4}. \eeq
\eitem
\vspace*{.2cm}

\noi 3. 
\vspace*{-1cm}

\beq
\begin{array}{|c|c|c||c|c|c||c|}
\hline
I_j & I_{j+1} & I_{j+2} & F_j & F_{j+1} & F_{j+2} & Amplitude \\
\hline
p & p+1 & p & - & - & - & - \\
\hline
\end{array}
\eeq

The terms in Eq.~\eqref{efbo2} take the state $(I_j,I_{j+1}, I_{j+2})$ to a 
state which is not relevant to the calculation of $H_{eff}$.
\vspace*{.2cm}

\noi 4.
\vspace*{-1cm}

\beq
\begin{array}{|c|c|c||c|c|c||c|}
\hline
I_j & I_{j+1} & I_{j+2} & F_j & F_{j+1} & F_{j+2} & Amplitude \\
\hline 
\hline
p & p+1 & p+1 & p & p+2 & p & \frac{i\ga TU}{2} ~\sqrt{p+1}\sqrt{p+2} \\
p & p+2 & p & p+1 & p+1 & p & - \frac{i\ga TU}{2} ~\sqrt{p+1}\sqrt{p+2} \\
\hline
\end{array}
\eeq
\vspace*{.2cm}

\bitem
\item The energy cost from the on-site energy is $-U$.
\item The total contribution is 
\beq -\frac{(p+1)(p+2) \ga^2 T^2 U}{4}. \eeq
\eitem
\vspace*{.2cm}

\noi 5. 
\vspace*{-1cm}

\beq
\begin{array}{|c|c|c||c|c|c||c|}
\hline
I_j & I_{j+1} & I_{j+2} & F_j & F_{j+1} & F_{j+2} & Amplitude \\
\hline 
\hline
p+1 & p & p & p+1 & p-1 & p+1 & \frac{i\ga TU}{2} ~\sqrt{p}\sqrt{p+1} \\
p+1 & p-1 & p+1 & p & p & p+1 & - \frac{i\ga TU}{2} ~\sqrt{p}\sqrt{p+1} \\
\hline
\end{array}
\eeq
\vspace*{.2cm}

\bitem
\item The energy denominator is $-U$.
\item The total contribution is 
\beq -\frac{p(p+1) \ga^2 T^2 U}{4}. \eeq
\eitem
\vspace*{.2cm}

\noi 6. 
\vspace*{-1cm}

\beq
\begin{array}{|c|c|c||c|c|c||c|}
\hline
I_j & I_{j+1} & I_{j+2} & F_j & F_{j+1} & F_{j+2} & Amplitude \\
\hline
p+1 & p & p+1 & - & - & - & - \\
\hline
\end{array}
\eeq
\vspace*{.2cm}

The terms in Eq.~\eqref{efbo2} take the state $(I_j,I_{j+1}, I_{j+2})$ to a 
state which is not relevant to the calculation of $H_{eff}$.
\vspace*{.2cm}

\noi 7.
\vspace*{-1cm}

\beq
\begin{array}{|c|c|c||c|c|c||c|}
\hline
I_j & I_{j+1} & I_{j+2} & F_j & F_{j+1} & F_{j+2} & Amplitude \\
\hline 
\hline
p+1 & p+1 & p & p & p+2 & p & \frac{i\ga TU}{2} ~\sqrt{p+1}\sqrt{p+2} \\
p & p+2 & p & p & p+1 & p+1 & - \frac{-i\ga TU}{2} ~\sqrt{p+1}\sqrt{p+2} \\
\hline
\end{array}
\eeq
\vspace*{.2cm}

\bitem
\item The energy denominator $-U$.
\item The total contribution is
\beq -\frac{(p+1)(p+2) \ga^2 T^2 U}{4}. \eeq
\eitem
\vspace*{.2cm}

\noi 8. 
\vspace*{-1cm}

\beq
\begin{array}{|c|c|c||c|c|c||c|}
\hline
I_j & I_{j+1} & I_{j+2} & F_j & F_{j+1} & F_{j+2} & Amplitude \\
\hline
p+1 & p+1 & p+1 & - & - & - & - \\
\hline
\end{array}
\eeq
\vspace*{.2cm}

The terms in Eq.~\eqref{efbo2} take the state $(I_j,I_{j+1}, I_{j+2})$ to a 
state which is not relevant to the calculation of $H_{eff}$.
\vspace*{.2cm}

Looking at the processes in items 2, 4, 5 and 7 above, we see that all of
them interchange $n_j$ and $n_{j+2}$ keeping $n_{j+1}$ unchanged. 
\vspace*{.2cm}

\noi {\bf Direct contributions from terms of order $\ga^2 T^2 U$ involving
three sites:}
\vspace*{.1cm}

\beq
\begin{array}{|c|c|c||c|c|c||c|}
\hline
I_j & I_{j+1} & I_{j+2} & F_j & F_{j+1} & F_{j+2} & Amplitude \\
\hline 
\hline
p & p & p+1 & p+1 & p & p & \frac{\ga^2 T^2 U}{3} ~p(p+1) \\
p+1 & p & p & p & p & p+1 & \frac{\ga^2 T^2 U}{6} ~p(p+1) \\
p & p+1 & p+1 & p+1 & p+1 & p & \frac{\ga^2 T^2 U}{6} ~(p+2)(p+1) \\
p+1 & p+1 & p & p & p+1 & p+1 & \frac{\ga^2 T^2 U}{6} ~(p+2)(p+1) \\
\hline
\end{array}
\eeq
\vspace*{.2cm}

We see that these processes also interchange $n_j$ and $n_{j+2}$ keeping 
$n_{j+1}$ unchanged. Adding up the contributions of the second order processes
and direct contributions involving three sites, we obtain the following table.
\vspace*{.2cm}

\beq
\begin{array}{|c|c|c||c|c|c||c|}
\hline
I_j & I_{j+1} & I_{j+2} & F_j & F_{j+1} & F_{j+2} & Amplitude \\
\hline 
\hline
p & p & p+1 & p+1 & p & p & (\frac{1}{3}-\frac{1}{4}) \ga^2 T^2 U ~p(p+1) = 
\frac{\ga^2 T^2 U}{12} ~p(p+1) \\
p+1 & p & p & p & p & p+1 & \frac{\ga^2 T^2 U}{12} ~p(p+1)\\
p & p+1 & p+1 & p+1 & p+1 & p & \frac{\ga^2 T^2 U}{12} ~(p+2)(p+1) \\
p+1 & p+1 & p & p & p+1 & p+1 & \frac{\ga^2 T^2 U}{12} ~(p+2)(p+1) \\
\hline
\end{array}
\label{table97} \eeq

We now recall from Eq.~\eqref{nj} that $n_j$ is related to the pseudo-spin 
$\si^z_j$. Hence the terms in \eqref{table97} can be fitted to a 
three-spin interaction of the form 
\beq (b_1 ~+~ b_2 \si^z_{j+1}) ~(\si^+_j \si^-_{j+2} ~+~ \si^-_j \si^+_{j+2}).
\eeq
To be explicit, we find that this part of $H_{eff}$ is given by
\beq \frac{\ga^2 T^2 U}{12} (p+1) (p+1+\si^z_{j+1})(\si^+_j \si^-_{j+2} + 
\si^-_j \si^+_{j+2}). \label{hlow2} \eeq
\vspace*{.2cm}

\noi {\bf Direct contributions from terms of order $\ga^2 T^2 U$ involving
two sites:}
\vspace*{.2cm}

Finally, we find that the terms in Eq.~\eqref{efbo3} contribute to terms
in $H_{eff}$ which involve only two sites. Using Eq.~\eqref{nj}, we find that
\bea && - \frac{\ga^2 T^2 U}{3} ~\sum_j ~\Big[ 2n_j(n_j-1) -4n_j n_{j+1} \Big]
\non \\ 
&=& \frac{\ga^2 T^2 U}{3} ~\sum_j ~\left[ 2 (p+1) \si^z_j + \si^z_j 
\si^z_{j+1} + (2p^2+4p+1) \right]. \label{hlow3} \eea

Putting together Eqs.~\eqref{hlow1}, \eqref{hlow2} and \eqref{hlow3}, 
we find the complete effective Hamiltonian shown in Eq.~\eqref{heffp} 
in the main text.

\end{document}